\documentclass[12pt,preprint]{aastex}

\def\deg{\ifmmode^\circ\else$^\circ$\fi}

\def\mic{~$\mu$m}

\def\et{{et al.~}}

\def\arcs{\ifmmode {''}\else $''$\fi}
\def\arcm{\ifmmode {'}\else $'$\fi}
\def\parcs{\sa=.07em \sb=.03em
     \ifmmode $\rlap{.}$^{\scriptscriptstyle\prime\kern -\sb\prime}$\kern -\sa$
     \else \rlap{.}$^{\scriptscriptstyle\prime\kern -\sb\prime}$\kern -\sa\fi}
\def\parcm{\sa=.08em \sb=.03em
     \ifmmode $\rlap{.}\kern\sa$^{\scriptscriptstyle\prime}$\kern-\sb$
     \else \rlap{.}\kern\sa$^{\scriptscriptstyle\prime}$\kern-\sb\fi}

\def\spose#1{\hbox to 0pt{#1\hss}}
\def\simlt{\mathrel{\spose{\lower 3pt\hbox{$\mathchar"218$}}
     \raise 2.0pt\hbox{$\mathchar"13C$}}}
\def\simgt{\mathrel{\spose{\lower 3pt\hbox{$\mathchar"218$}}
     \raise 2.0pt\hbox{$\mathchar"13E$}}}
\def\lsim{\rlap{$<$}{\lower 1.0ex\hbox{$\sim$}}}
\def\gsim{\rlap{$>$}{\lower 1.0ex\hbox{$\sim$}}}

\begin{document}


\title{1--1.4 Micron Spectral Atlas of Stars}

\author{M. A. Malkan \altaffilmark{1}, E. K. Hicks \altaffilmark{1}, H. I. Teplitz\altaffilmark{2,}\altaffilmark{3}, I. M. McLean \altaffilmark{1}, H. Sugai\altaffilmark{4}, J. Guichard\altaffilmark{5}}

\altaffiltext{1}{Department of Physics and Astronomy, University of California,   Los Angeles, CA, 90095-1562 }
\altaffiltext{2}{NOAO Research Associate} 
\altaffiltext{3}{Laboratory for Astronomy and Solar Physics, Code 681, Goddard Space Flight Center, Greenbelt MD 20771 Electronic mail: hit@binary.gsfc.nasa.gov}
\altaffiltext{4}{Department of Astronomy, Kyoto University, Sakyo-ku, Kyoto 606-8502 Japan}
\altaffiltext{5}{Instituto Nacional de Astrofisica, Optica y Electronica, 20 Luis Enrique Erro 1, Tonantzintla, Puebla, 72840, Mexico}

\begin{abstract}

  We present a catalog of J-band (1.08\mic~to 1.35\mic) stellar
  spectra at low resolution (R $\sim$ 400).  The targets consist of 105
  stars ranging in spectral type from O9.5 to M7 and luminosity
  classes I through V.  The relatively featureless spectra of hot
  stars, earlier than A4, can be used to remove the atmospheric
  features which dominate ground-based J-band spectroscopy.  We
  measure equivalent widths for three absorption lines and nine
  blended features which we identify in the spectra.  Using detailed
  comparison with higher resolution spectra, we demonstrate that low
  resolution data can be used for stellar classification, since
  several features depend on the effective temperature and gravity.
  For example The CN index (1.096 - 1.104\mic) decreases with
  temperature, but the strength of a blended feature at 1.28\mic~
  (consisting of primarily P$\beta$) increases.  The slope of a star's
  spectrum can also be used to estimate its effective temperature.
  The luminosity class of a star correlates with the ratio of the Mg I
  (1.1831\mic) line to a blend of several species at 1.16\mic.  Using
  these indicators, a star can be classified to within several
  subclasses.  Fifteen stars with particularly high and low metal
  abundances are included in the catalog and some spectral dependence
  on metal abundance is also found.  

\end{abstract}


\section{Introduction}
The widespread availability of infrared array detectors, coupled with
a new generation of infrared spectrometers (e.g. McLean 1994; McLean
et al. 1998) have allowed for increasingly routine spectroscopy of
astrophysically important stellar absorption lines.  Near-infrared
(NIR) spectroscopy is essential to the study of young and low mass
stellar objects (e.g.McLean et al.\ 2000), as well as regions of the galaxy suffering large extinction (e.g. Figer, McLean and Morris 1999).  NIR absorption 
lines are also being observed in composite stellar systems, 
such as star clusters and local galaxies (Mouhcine \& Lancon 2001).  Observations of the rest-frame NIR spectra of galaxies at appreciable redshift will
soon be possible with SIRTF (Fanson et al. 1998) and NGST (Stockman 1997).  
There is consequently a growing need for accurate modeling of the NIR spectral
range, which in turn must rely on NIR observations of all types of
stars.

Several excellent atlases of stellar spectra have been published for
the H (1.5--1.8\mic) and K (2.0--2.4\mic) atmospheric windows (Lancon
$\&$ Rocca 1992; Wallace $\&$ Hinkle 1996). These data have shown that
stellar atmosphere models do not {\it a priori}~predict acceptable
spectral energy distributions (SED's) in the near-IR, but must be refined 
based on observations.  Published spectra in the J (1.05--1.35\mic) window 
are limited to a few specialized studies (e.g. Jones
et al.\ 1996) and the recent atlas of Wallace et al.\ (2000).  Though 
stellar spectral classification is easiest to do with high resolution 
data, as obtained in the above mentioned studies, lower resolution is 
necessary for observations of a substantial number of objects (e.g. 
survey of distant galaxies).

In this paper, we present the results of a spectroscopic survey of
normal stars in the J-band.  Using a low resolution grism
we observed 105 stars in the northern hemisphere.  The sample
spans spectral types O9.5-M7 and luminosity classes I-V.  Some
preliminary results were presented in Hicks \et (2000).  The observations were
taken over the course of several years, often as a ``backup'' program
in non-photometric weather.  As we outline below, care has been taken
to determine the best correction for the atmospheric absorption.
The observations are described in $\S$ 2, and $\S$ 3 discusses
the data reduction techniques.  One initial impetus for the inquiry
was the need for good atmospheric standards for J-band spectroscopy of
extragalactic targets, and we discuss the results of that
investigation in $\S$ 3, as well.  In $\S$ 4 we provide the
atlas, line lists and discuss line identification.  Section 5 presents a
comparison with previous studies at higher resolution and $\S$ 6
discusses the trends observed in the data.  A
near-infrared stellar spectral classification scheme is discussed in
$\S$ 7.

\section{Observations}

The UCLA 2-Channel spectrograph (McLean et al 1994; known colloquially
as GEMINI for its twin detectors) has been used regularly during
bright runs over the last 8 years on the Shane 3.0-m reflector at Lick
Observatory.  During this time we have obtained a large archive of
J-band stellar spectroscopy, which are presented here as an Atlas for
future applications.  The Atlas has been assembled during an ongoing 
``backup'' program during nights of
non-photometric conditions.  Supplemental observations for the Atlas
were made at the start and the end of a night's observing in good
weather.  Additional J-band spectra were obtained to correct for
atmospheric absorption in other programs.

The Atlas contains 105 stars (see Table 1), 99 selected from the Bright Star Catalog (1982) and 6 from the Henry Draper Catalog (1995).  Table 1 includes the catalog name, the spectral class, effective temperature, and, when available, the metallicity.  Fifty-eight are main sequence stars and 46 are giants or supergiants.  Fifteen stars were chosen because high-resolution optical spectroscopy has shown that they have unusually high or low heavy-element abundances (Cayrel de Strobel \et 1997).

All observations were made using the Rockwell HgCdTe 256 x 256 array
detector and grism spectroscopy in the J-band.  One pixel corresponds
to about 11.6 \AA, and the 2-pixel (1.5 arcseconds) wide slit
yielded a typical spectral resolution of $R\sim 400$.  Each star was
observed consecutively in two positions, separated by 20--40
arcseconds along the North/South slit.  The two images were used as
the sky frames for each other, by direct subtraction.  In many cases
no telescope guiding was attempted during the short ($\le 60$~second)
exposures.  This sometimes resulted in the star drifting partly out of
the slit during the second exposure.  If the signal dropped by more
than a factor of two, the star was recentered on the slit for a third
exposure.  In any case, the absolute flux levels are not of interest
in this study.  In general the spectral shape (i.e. relative fluxes
from 1.05 to 1.35\mic) is accurately measured.

\section{Data Reduction}
All spectra were initially processed with IRAF\footnote[1]{IRAF is
  distributed by NOAO, which is operated by AURA Inc., under cooperative 
  agreement with the NSF} scripts based on the APEXTRACT package.  After
eliminating bad pixels, images were sky-subtracted and divided by
a normalized flat field.  The flat frame was the difference between 
exposures of the dome ceiling with and without an illuminating lamp.
Next the top and bottom spectra were mapped, rectified, and extracted,
using a 7$\arcs$-long slit region, with local sky subtraction.  The
arithmetic average of the two spectra was obtained.

Near-IR spectroscopy, even at low resolution, requires careful sky
subtraction and correction for atmospheric absorption.  The strength
of OH radical emission in the night sky varies rapidly with time,
requiring contemporaneous sky measurements.  Wavelength calibration is
complicated by the density of night sky lines, which are often
unresolved at low resolution; good calibration data from comparison 
lamps is thus required.  The spectra were assigned a wavelength scale,
usually obtained from a Chebyshev polynomial fit to 11--13 isolated
arc lines in the spectra of a comparison lamp of Kr, Ar, or Ar with
Ne.  The GEMINI spectrograph has been shown to be stable (see Sugai
\et 1997) allowing the use of calibration data taken at the end of the
night or even the following night.  In a few cases the wavelength
calibrations were obtained after a longer gap in time during the
observing run.  As a result, wavelength shifts of up to 1 or 2 pixels
(11.6 \AA\ to 23.2 \AA) sometimes occurred.  In the worst case, the
wavelength scales of these stellar spectra were re-calibrated using
the strongest OH night-sky lines visible in the stellar spectra
themselves.  The resulting wavelengths, using either a comparison lamp 
or sky lines, are accurate to better than one
pixel, with $\sigma$($\lambda$) $\sim$ 8 \AA, except where indicated in Table 3.  Data taken during the
September 1999 run have less accurate wavelength scales because the grism 
wheel motions were not completely repeatable, resulting in an increased
uncertainty in the calibration of 15 \AA.  Typical examples of arcs and night-sky
lines are shown in Figure 1.  The measured full width at half-max (FWHM) 
of arc lines and night-sky lines is 2.4 to 2.8 pixels (28 to 32 \AA).

The final step in the reduction is the atmospheric absorption
correction.  Stellar spectra were divided by a telluric extinction
template spectrum (see Figure 2).  Stars hotter than A4 (8550 K) have
only two detectable absorption features (P$\gamma$~and P$\beta$), and
so may be used to create atmospheric templates.  After interpolating
across these features what remains is purely atmospheric absorption.
We used a uniform template to correct all the stellar data (with the
exception of the September 1999 run, which was calibrated separately).
This template spectrum consists of two parts: the average of six
spectra of A1V stars in the wavelength interval 1.15--1.35\mic\ 
together with the average of one A3V and two A1V stars for the
1.07--1.10\mic\ region.  The September 1999 run data were corrected
for telluric absorption using the average of three A0 and two A2 stars
taken during that run.  In all of the stellar data, the worst region
of atmospheric absorption, from 1.10--1.15\mic, could not be fully
corrected.

Three other atmospheric absorption template spectra were used to confirm the
validity of our results.  The first check used the average of seven
A3V stars from our own data.  We also compared our results to those
obtained when using the absorption templates from Wallace \et
(1996) and Hinkle \& Wallace (1995); those authors created high
dispersion absorption templates in order to correct their observations
of the Sun (spectral type G2V) and of Arcturus (spectral type K2III), 
respectively.  Each of these atmospheric 
template spectra yield results similar to those given by the template 
originally used, confirming that our choice of templates removes the 
atmospheric absorption adequately.

The spectra in the Atlas have excellent  photon statistics, with over
ten thousand photons detected per pixel.  The repeatability of the
spectra measured at the top and bottom slit positions confirms that
errors due to sky subtraction and flat fielding are also below 1\%,
except in the 1.10--1.15\mic\ atmospheric absorption trough.  The
typical signal to noise ratio (SNR) in each spectrum is better than
100 per 2.5-pixel (30 \AA) resolution element.  Thus, absorption 
lines down to 1 \AA~equivalent width (or 2 \AA~in the most
uncertain part of the spectrum which is heavily affected by
atmospheric absorption) can be measured.

The data reduction process altered the intrinsic slope of the stellar spectra, though in a predictable way.  The steps in our reduction process that could alter the slope are the division by a flat field and the division by the atmospheric template.  With each run, flat fields with a slightly different wavelength dependence were used.  Since the wavelength dependence was not removed, this caused the spectra from different observing runs to have slopes that differ by as much as a factor of three for stars of a similar spectral type (which are expected to have similar spectral slopes).  These spectra were then divided by an atmospheric absorption A star template that was created from an average of several A stars from different runs.  Had the spectra been divided by a template created from only A stars observed on the same run, and thus corrected with the same flat field, then stars of the same spectral type from all runs would have similar slopes.  However, to create a high quality atmospheric template it was necessary to average A stars from several different observing runs.  The resulting slopes of the spectra were thus dependent on the run in which they were observed.  The dependence was removed by determining the correction necessary to force A1V stars to have a slope of zero,  which would be expected after dividing by an atmospheric template made from A1V stars.  This correction was determined for each run separately.  The resulting slope is then a differential slope that is the ratio of the star's slope to that of the A1V star atmospheric template.  Since we are only concerned with the slope of the spectra longward of 1.15\mic, this correction has only been applied to the spectra that were divided by the A1V atmospheric template.

All of the reduced spectra are publicly available in electronic
form\footnote[2]{{\tt http://www.astro.ucla.edu/$\sim$malkan/jspec/}}.  
Separate spectra are given for wavelengths shorter than 1.15\mic\ 
(``HR\#a.fits") and longward of 1.15\mic\ (``HR\#b.fits").

\section{Absorption Feature Identifications}

We identified three absorption lines and nine blends in our stellar spectra (see Table 2).  Each feature was required to satisfy the following criteria: 
\begin{enumerate}
\item The line must be present in the spectra of many different stars within a range of surface temperatures.
\item The central wavelength must be constant in all of those spectra,
to within our accuracy of one pixel $= \pm 11.6$ \AA.
\item The central wavelength of the line must agree with a known transition in the appropriate stellar atmosphere model.
\item The observed line widths must be reasonable (i.e. FWHM=30--40 \AA~in our spectra).
\item The presence and shape of the line must not depend on the choice of atmospheric extinction template used to correct for telluric absorption.
\end{enumerate}

The line identifications were based on comparison with high-resolution
spectra of the Sun (Wallace et al. 1996) and the K2III star Arcturus
(Hinkle and Wallace 1995).  In most cases, a spectral feature in our
data appears to be produced by the blending of 2 or even 3 strong
lines within one of our resolution elements.  The individual species
creating each blend, along with their absorption feature wavelengths, 
are listed in columns 2 and 3 of Table 2.   Where possible we have
estimated which line dominates by matching the average central wavelength
observed in our spectrum with the wavelength of the strongest line
of the blend in the solar or Arcturus spectra.  The central wavelength 
of the blend is given in column 4.  If a feature is not a blend, but 
composed of a single species, then the wavelength of this single 
absorption feature is given in column 4.  Often a change in the dominating 
species occurs at lower temperatures (in K and M stars) causing a shift 
in the blend wavelength.  This new central 
wavelength of the blend for the cooler stars is given in parenthesis in 
column 4.  Most lines are from neutral metals, with molecular features 
appearing in only the cooler stars.  Notes on the species in each blend 
for the hotter stars are given in column 5, and any changes in the 
species in a blend at cooler temperatures (in K and M stars) are given in column 6.  Figures 3a-e show blowups of the spectral regions with the line blends in the solar spectrum, Arcturus, and in the averages of several stars in each spectral class.  As expected, the only lines present in stars earlier than A4 were P$\gamma$~and P$\beta$ (called P$\gamma$~ and 1.28\mic~respectively in Table 2).

Equivalent widths for each of the features measured are given in Table 3.  Equivalent widths were measured with Gaussian fitting, and checked with the simpler direct-integration method between two ``continuum" points in the IRAF SPLOT routine.  Continuum points were placed along the apparent local continuum on either side of the absorption feature.  The agreement between the two methods was within 1$\%$, and the results for the Gaussian fitting are reported.  Both the central wavelength and the width of the feature were left as free parameters, except in the case of the 1.33\mic~blend (see below).  The uncertainty in the equivalent width is typically $\sigma$(EW) $\simeq$ 0.4 \AA, and, as stated, the uncertainty in the wavelength is $\sigma$($\lambda$) $\sim$ 8 \AA, except where noted.  In the cooler stars, the uncertainties for some of the blend equivalent width measurements are larger ($\sigma$(EW) = 0.8 \AA) due to the difficulty in estimating the true continuum level in a low resolution spectrum cut up by many absorption lines.  Most of the lines, even the blends, are unresolved, having FWHM consistent with the instrumental resolution (28 to 32 \AA).  The FWHM of the P$\gamma$~and the 1.28\mic~blend (which is composed primarily of P$\beta$) are occasionally found to be $\sim40$ \AA~because the lines are marginally resolved.  Measurements where the equivalent width measurement is particularly uncertain (where $\sigma(EW) > 0.4$ \AA) due to the difficulty in determining the continuum level or a FWHM greater than the instrumental resolution are indicated with a colon in Table 3.  The FWHMs of the 1.21\mic~and 1.33\mic~blends are greater than the instrumental resolution because their components are separated by 28 \AA~and 39 \AA, respectively.  In order to provide a more reliable measurement for the 1.33\mic~blend, the FWHM was fixed at 35 \AA~and the core depth was measured instead of the equivalent width.  The CN 0-0 index was measured from 10960 \AA~to 11040 \AA, with continuum measured in windows on either side at 10925-10960 \AA~and 11040-11100 \AA.   This choice of wavebands is a compromise based on the need to measure as much of the absorption as possible without including noisier points at the red and blue ends.  At most fifteen percent of the absorption was missed.

\section{Comparison to Previous Studies at Higher Resolution}

After we had completed our equivalent width measurements, Wallace et al. (2000, hereafter W2000) published results of a J-band spectroscopic survey of 88 MK standard stars, none coinciding with our sample of 105.  They had a higher resolution of $R\sim 3000$, and thus did not measure all of the same blends that we identified.  Many of the absorption features making up our blends could be measured individually at their higher resolution, or in some cases the individual features were not very strong and therefore were not measured by W2000.  

In order to compare our results with W2000, we degraded their digital spectra to our resolution and then measured the strongest features using our method for determining equivalent widths.  As expected, the 12 features we identified in our spectra are also the strongest features in their spectra.  Our measurements of these 12 features in their degraded spectra are plotted along with our data in the next section (small black circles Figure 5).  The equivalent widths we measured in the degraded W2000 spectra are similar to those in our own data, and the same dependences on the effective temperature are seen.  The W2000 measurements have less scatter due to the higher original resolution.


A few features (CN 0-0 index, 1.16\mic, Mg I, and 1.28\mic) were measured in both the high and low resolution spectra, making it possible to relate the measurements published by W2000 directly to what we measured.  Two of the features, Mg I and 1.28\mic~(known as Mg I and H P$\beta$ in W2000) are exactly the same.  Our 1.16\mic~blend overlaps their Fe I feature, and we both measure the same Fe I lines.  Our CN index is also contained within the CN window W2000 measured.  For the 1.16\mic, Mg I, and 1.28\mic~features, the equivalent widths we measured at low resolution using our technique were plotted against the high resolution measurements published by W2000 (see Figures 4a-c).  It was found that the low resolution measurements have equivalent widths about 2 times the equivalent widths measured at high resolution.  This increase in  width at lower resolution is due to additional features included in the wider window we measured compared to that measured by W2000 in the case of the Mg I and 1.28\mic~features.  For the 1.16\mic~feature, our window has comparable width to the W2000 feature.  However, the slight displacement in wavelength causes our window to include additional features, resulting in an increased equivalent width at lower resolution.  The transformations found for each of the blends are as follows, in units of angstroms:
\begin{equation}
\eqnum{1.1}
EW(1.16)_{low} = 2.22 * EW(1.16)_{high} - 0.67
\end{equation}
\begin{equation}
\eqnum{1.2}
EW(1.16)_{low} = 1.58 * EW(1.16)_{high}
\end{equation}
\begin{equation}
\eqnum{1.3}
EW(Mg~I)_{low} = 1.95 * EW(Mg~I)_{high} + 0.43	
\end{equation}
\begin{equation}
\eqnum{1.4}
EW(Mg~I)_{low} = 2.57 * EW(Mg~I)_{high} 	
\end{equation}
\begin{equation}
\eqnum{1.5}
EW(1.28)_{low} = 2.16 * EW(1.28)_{high}	  
\end{equation}   
The first equation given for 1.16\mic~and for Mg I is the best fit line with no constraint on the y-axis intercept, and the second equation is the best fit line with a forced y-intercept of zero.  For P$\beta$, the fits with and without forcing the intercept to zero were similar enough to just use the forced intercept fit.  The linear Pearson's correlations coefficient of these transformation relations are 0.254, 0.3484, and 0.9442 for equations 1.1, 1.3, and 1.5, respectively.

Jones \et (1996) obtained 1.16--1.22\mic\ intermediate-resolution spectra of 9 M dwarfs with spectral types from M2 to M6.  Where our lower-resolution observations overlap with theirs, there is reasonable agreement.  For example, we both find equivalent widths of the 1.17\mic\ blend (dominated by K I in M stars) which {\it increase} from 1.5 to 4 \AA\ as the effective temperature drops from roughly 4000 to 3000 K.  Our observations of the coolest M stars also appear consistent with a {\it decrease} from 2 to 1 \AA\ in the equivalent width of Mg I absorption (at 1.183\mic). Although Jones \et did not measure the Si I and 1.21\mic\ absorptions, they are clearly visible in their figure 1. Their spectra again confirm our observations which show that both of these absorptions drop sharply in strength at $T_{eff} \le $3500 K.  In these cool stars Fe I absorption should dominate both the 1.19\mic~and 1.20\mic~blends (see figure 3 of Jones \et).  Our M star observations are consistent with average equivalent widths of 1.5--2.0 \AA\ for both of these.  These line strengths agree well with the stellar atmosphere {\it models} in Jones et al., although their observations are systematically lower (around 1 \AA\ or less).  We could assume that the model predictions are correct.  Alternatively, we could seek an explanation for why our observed equivalent widths appear higher than those of Jones \et\  The fact that most of our M stars are giants, not dwarfs, could perhaps explain some of this possible discrepancy (see the Jones \et models for varying gravity).  And broad FeH absorption also inflates our measurements of the 1.20\mic\ blend in the coolest stars.  The other possibility is that our equivalent width measurements are driven systematically higher by having 2.5 times lower spectral resolution than Jones et al.

\section{Blend Analysis}

\subsection{Dependence on Effective Temperature}

Several of the blends identified exhibit a dependence on effective temperature, and thus on spectral type.  At different stellar surface temperatures, the dominating transitions that make up the blend change relative strengths, or an entirely new line may appear.  The species contributing to each blend for both the hotter (T$_{eff} \ge$ 5000 K) and cooler stars (T$_{eff} <$ 5000 K) are described in Table 2.  This change in dominating species can cause the width and/or central wavelength of the blend to shift with temperature.  The equivalent widths and FWHM of most of the blends (except for the Mg I, Si I, and 1.19\mic~features) increase in the cooler stars.  Plots of the equivalent width versus the effective temperature for all identified features are shown in Figures 5a-l.  The 1.19\mic, 1.20\mic, 1.21\mic, 1.28\mic, and 1.33\mic~blends all show systematic shifts in wavelength as well as in equivalent width (with the exception of the 1.19\mic~feature which does not exhibit a change in equivalent width).  The two best examples of a shift in central wavelength are shown in Figures 6a and 6b, which plots central wavelength versus the effective temperature.  For most of the blends the change in central wavelength and/or equivalent width and FWHM is due to an introduction of Ti I or CN 0-0 absorption, or a change in the relative strength of the original dominating species.  Most commonly, CN 0-0 and Ti I features strengthen and C I absorption decreases as the temperature decreases.  The details of the dependence of absorption on effective temperature for the individual features follows.

{\it P$\gamma$} -  This hydrogen absorption feature was measured in stars with temperatures greater than 4500 K (G9).  The window measured was from 10890 \AA~to 10990 \AA.  The feature peaks in equivalent width (see Figure 5a) around 10 \AA~at a temperature of 10,000 K (A0V stars).  In cooler stars the equivalent width drops significantly, to around 2 \AA~at 4500 K.  For stars cooler than 4500 K, P$\gamma$ is very weak and is lost in a strong blend of CN 0-0 absorption which will be discussed below.  There are only four stars hotter than 10,000 K with confident measurements, and each has a lower equivalent width than the peak. 

{\it CN 0-0 index} - A strong trough of CN 0-0 absorption is present at the blue edge of our spectra in the coolest stars (T$_{eff} <$ 4500 K).  The window measured was from 10960 \AA~to 11040 \AA.  The CN 0-0 feature extends further to the blue, beyond the reach of our spectra.  This results in our measurement underestimating the band strength, but not by more than 15$\%$.  Figure 5b shows the dependence of the equivalent width on temperature, with an increase in equivalent width with decreasing temperature.

{\it 1.16\mic~blend} - This feature is dominated by Si I, Fe I, and Cr I equally at the hotter temperatures, but as the temperature decreases Fe becomes more dominant and CN 0-0 is introduced.  As a result of the introduction of the CN 0-0 features throughout the blend, the equivalent width increases at lower temperatures (see Figure 5c).  This feature was measured from 11570 \AA~to 11630 \AA.  

{\it 1.17\mic~blend} - Measured from 11725 \AA~to 11800 \AA, this feature is a blend of C I, K I, and Fe I  in hotter stars.  At cooler temperatures the C I lines weaken and CN 0-0 is introduced.  In the coolest stars (M stars), K I (1.177\mic) dominates, with strong K I absorption at 1.169\mic~also visible.  A slight increase in equivalent width with decreasing temperature is seen in Figure 5d.

{\it Mg I feature} - This feature consists of a single Mg I line and was measured from 11800 \AA~to 11865 \AA.  Since the feature has no dependence on temperature and is isolated at all temperatures, no change in equivalent width or central wavelength is seen (see Figure 5e).

{\it 1.19\mic~blend} - Consisting primarily of two Fe I lines and some weaker C I lines at hotter temperatures, this line was measured from 11850 \AA~to 11910 \AA.  At cooler temperatures the C I absorption weakens and a Ti I line at 1.1896\mic~strengthens along with several CN 0-0 features throughout the blend.  The introduction of the Ti I absorption causes the central wavelength of the feature to change from 1.189\mic~to 1.188\mic~in cooler stars.  Any dependence of the equivalent width on temperature is not detectable in our data (see Figure 5f).  

{\it 1.20\mic~blend} - This is a blend of two Si I lines and a Fe I line in hotter stars, with the Si I line at 11987 \AA~the strongest of the three.  This feature was measured from 11960 \AA~to 12000 \AA.  At cooler temperatures a Ti I line at 1.1977\mic~is present, causing not only an increase in the width of the line at cooler temperatures (see Figure 5g), but also the wavelength shift from 1.199\mic~to 1.198\mic~seen in Figure 6a.  In the coolest M stars, this feature changes further to be dominated by a broad trough of FeH absorption (Jones \et 1996).  Our measurements for stars with $T_{eff} \le $ 3800 K are consistent with the 1.5 \AA\ values observed in and predicted for M dwarfs by Jones \et (1996).  

{\it Si I feature} - Measured from 12015 \AA~to 12070 \AA, this feature consists of a single Si I absorption line.  With no contamination at cooler temperatures, no change in equivalent width (see Figure 5h) or central wavelength with temperature is detected in this feature.

{\it 1.21\mic~blend} - This feature was measured from 12070 \AA~to 12150 \AA.  It is a blend of several Si I lines and a Mg I line.  The central wavelength of the feature is shifted from 1.210\mic~to 1.209\mic~in cooler stars as the reddest two Si I lines weaken, while the bluer Si I and Mg I lines remain strong.  CN 0-0 is also introduced throughout the blend, with the CN 0-0 on the bluer side slightly stronger than that on the redder side.  This unevenness of the CN 0-0 also contributes to the shift in wavelength of the blend, as well as causing the equivalent width to increase at lower temperatures (see Figure 5i).

{\it 1.28\mic~blend} - This blend is dominated by P$\beta$ at 12822 \AA~in hotter stars.  At lower temperatures a Ti I line at 12825 \AA~increases in strength while the hydrogen line decreases, causing the Ti I line to be the dominating species.  The feature was measured from 12760\ AA~to 12890 \AA~in stars of all temperatures.  Figure 5j shows that in hotter stars this feature acts very much like the other hydrogen feature measured, P$\gamma$, with a peak in equivalent width around 10 \AA~at a temperature of 10,000 K (A0V stars).  In cooler stars the equivalent width deceases to around 2 \AA, while for the four stars with temperatures hotter than 10,000 K it is around 5 \AA.  A wavelength shift is also measured in this blend (see Figure 6b).  For K and M stars the wavelength shifts to the red because of the Ti I absorption feature, causing the wavelength to shift from 1.282\mic~for hotter stars to 1.283\mic~in the cooler stars. 

{\it 1.31\mic~blend} - This blend consists of Si I and Al I lines and was measured from 13080 \AA~to 13200 \AA.  The Al I line at 13126 \AA~and the Al I line at 13180 \AA~are the dominating lines.  Figure 5k shows a slight increase in equivalent width of this feature with decreasing temperature.

{\it 1.33\mic~blend} - This blend is made up of several species with Si I and Fe I lines dominating.  The window in which this feature was measured is from 13250 \AA~to 13335 \AA.  In cooler stars the Fe I and Ca I lines weaken and the Mn I lines strengthen.  This change in dominating species causes a wavelength shift from 1.332\mic~to 1.331\mic~in cooler stars.  At the lowest stellar temperatures, a band of $H_2O$ absorption (Jones \et 1996) cuts into the red side of this blend, making it very difficult to measure.  At cooler temperatures the equivalent width of this feature increases slightly (see Figure 5l).

\subsection{Dependence on Gravity}

Along with a dependence on the effective temperature, several of the blends identified have a dependence on the luminosity class of the star and thus on the gravity.  This is most noticeable in the F through K spectral classes where we have reasonably large samples of dwarfs, giants and supergiants.  The strongest case is the Mg I line which has a larger equivalent width for dwarfs (luminosity classes IV and V) than for giants (luminosity classes I, II, and III) at a given effective temperature (see Figure 5e).  The 1.16\mic~blend exhibits the opposite dependence on gravity, having a larger equivalent width for giants (see Figure 5c).  1.31\mic~has the same dependence, but to a lesser degree.  None of the other blends show much of a dependence on gravity.
   
\subsection{Dependence on Metallicity}

Fifteen of the stars in our sample are known from high resolution optical spectroscopy to have high or low metal abundances (see logarithmic metallicity values in the last column of Table 1).  The high and low metallicity stars are shown in Figures 5 and 6, with high abundance stars defined as $[Fe/H] >  0.0$ and low abundance as $[Fe/H] < -0.3$.  For a given temperature, those stars with high (low) metallicity tend to have enhanced (diminished) equivalent width for some metal absorption blends, relative to the majority of normal stars (presumed to have solar, Population I abundances).  The strongest case is the 1.19\mic~blend where $\Delta$EW/$\Delta$log[Fe/H] $\sim$ 2.  The 1.28\mic~blend also has a positive dependence on metallicity, which is surprising, since the feature is dominated by hydrogen.  The dependence might be due to the introduction of Ti I absorption at lower temperatures.  Some of the blends (P$\gamma$, 1.17\mic, 1.20\mic, and 1.21\mic) have no apparent dependence on metallicity within the accuracy of our measurements.

\section{IR Stellar Spectral Classification}

\subsection{Temperature Indicator}
The blends most sensitive to the effective temperature are 1.28\mic~and 1.31\mic, as well as the CN index.   Hotter stars have higher 1.28\mic~equivalent widths, while  1.31\mic~and the CN index have the opposite temperature dependence.  Using these relationships between equivalent width and effective temperature, it is possible to estimate a given star's effective temperature based on a measurement of the equivalent width of one or two absorption features.  The dependence of 1.31\mic~is not as strong as that of 1.28\mic.  However, 1.28\mic~is only useful for effective temperatures between 5500 K and 9500 K, due to the addition of the Ti I line at these lower temperatures, and the turndown in the equivalent width for spectral types earlier than A0.  Since the CN 0-0 absorption only occurs in the coolest stars, the CN index is useful as a temperature indicator for stars with effective temperatures less than 5500 K.  If CN 0-0 absorption is detected,  the star has an effective temperature less than 5500 K;  if it is absent then the 1.28\mic~blend can be measured, and a temperature greater than 5500 K will be found.  Using the equivalent width of the 1.28\mic~blend it is possible to estimate an effective temperature to within 1600 K, while using the CN index will give an effective temperature to within 1300 K.  To quantify these dependencies on effective temperature the following relationships are used:
\begin{equation}
\eqnum{2}
T(1.28) = 1139 * EW(1.28) + 2211
\end{equation}
\begin{equation}
\eqnum{3}
T(CN) = -677 * EW(CN) + 5905
\end{equation}
Figure 7a shows the 1.28\mic~blend equivalent width versus the published temperatures for all stars with 5500 $<$ T $<$ 9500 K, except those stars with uncertain equivalent width measurements ($\sigma$(EW) $>$ 0.4 \AA) or high/low metallicity (see $\S$ 6.3), and the fit given by Equation 2.  The same plot of the CN 0-0 index equivalent width for stars with T $<$ 6000 K is given with Figure 7b, with the fit given by Equation 3 shown.

Another method to determine the effective temperature of a star is to use the spectral shape.  As explained in $\S$3, the slope of the final stellar spectrum is a result of the slope of the star divided by the slope of the A star atmospheric template.  The slope of the spectra are found to be dependent on the effective temperature with a greater, positive (with respect to wavelength) slope for stars with cooler effective temperatures.  For an A star, which intrinsically has a slope close to that of a 10,000 K black body, this {\it differential} slope is then measured to be about zero, while for stars with cooler effective temperatures the slope increases.  The slope of these stellar spectra ratios were determined by fitting a power law ($F_{\lambda} = C\lambda^{\alpha}$) to the final ratio spectrum between 1.14\mic~to 1.34\mic.  It was found that the differential slope $\alpha_{diff}$ depends on the effective temperature in the following way (see Figure 8):
\begin{equation}
\eqnum{4}
T(slope) = -2325 * \alpha_{diff} + 9574
\end{equation}
\noindent This relationship holds for temperatures less than 13500 K.  
Using the slope of the ratio of the stellar spectrum to the A star atmospheric template, it is possible to determine the effective temperature of a star to within 2500 K.  The temperature determined from the differential slope of the spectrum agrees with the temperature given by the equivalent width to within 2000 K (see Figure 9).  

\subsection{Surface Gravity Indicator}
As discussed earlier, Mg I, 1.16\mic, and P$\gamma$ are particularly sensitive to the luminosity class of the star.  To quantify this dependence, the following equation was used to make a fit of one blend plotted against another:
\begin{equation}
\eqnum{5}
EW(1st~blend) = constant1 + constant2 * EW(2nd~blend) + constant3 * Lum.~Class
\end{equation}
\noindent where the luminosity class is represented by I=1, II=2, etc.  The form of Equation 5 is not necessarily the best representation of the relationship between luminosity class and equivalent width, but was chosen for its simplicity.  The dominant term, as expected, is the equivalent width of the 2nd line, but for those lines which have a strong dependence on the surface gravity the third term is significant.  Those stars with known high or low metallicities were not included in the fit, nor were stars with uncertain equivalent width measurements ($\sigma$(EW) $>$ 0.4 \AA).  When fitting the equivalent width of Mg I versus the equivalent width of any of the other blends, the Mg I luminosity class coefficient is large and positive, indicating that dwarfs have higher equivalent widths than giants of similar effective temperature.  A similar result was found for P$\gamma$, but the coefficient is smaller than the one found for Mg I.  1.16\mic~has the opposite characteristic, having a large, negative coefficient multiplying the luminosity class in Equation 4.  This indicates that 1.16\mic~is a good indicator of giants, having a higher equivalent width for giants than it does for dwarfs.  The ratio of the equivalent width of Mg I to the equivalent width of the 1.16\mic~blend can therefore be used as an indicator of the luminosity to within one luminosity class, as illustrated in Figure 10.  The greater the ratio, the lower the luminosity class (see Figure 10), or, quantitatively stated:
\begin{equation}
\eqnum{6} 
EW(1.16) = 2.696 * EW(Mg~I) - 0.73 * Lum.~Class 
\end{equation}
This fit was done with all luminosity classes weighted equally.  Thus, the above relationship is not the best fit to each luminosity class individually, but rather to an average of the luminosity classes.   

\subsection{Classification Scheme}

As has been illustrated above, the ratio of the equivalent widths of Mg I to 1.16\mic~is an indicator of the luminosity class, while the equivalent width of either 1.28\mic~or the CN index can be used as an indicator of the effective temperature.  Also, the slope of the continuum can be used to estimate the effective temperature.  Combining this information, a classification can be made of a star to within three subclasses based on easily measured quantities.  Of course, the slope of a star is not completely reddening independent.  Using the temperature measurement based on the equivalent width of the 1.28\mic~bland or CN index one could use the observed slope to deduce independently the approximate reddening of a given star.

\section{Conclusions}

The principal result of our observations of hot stars is that, at least at our spectral resolution, it is safe to approximate their spectra as featureless blackbodies, after removing the P$\beta$~and P$\gamma$~absorption lines.  Thus these stars are useful for measuring the atmospheric extinction.

Even at the low resolution of this study, nine blends and three lines were identified.  As expected, the most important stellar property controlling the absorption line strengths we measured is the effective stellar surface temperature.  The strength of the Paschen lines peaks at a maximum equivalent width of 10 \AA~at A0 and declines at higher and lower temperatures.  The neutral metal lines are not strong until later than A7, and their strength increases in cooler stars.  The most noticeable change in the spectra as the temperature decreases is the increase of CN 0-0 and Ti I lines and the relative decrease of C lines.  There is, however, significant cosmic scatter (i.e., larger than observational) in the line strengths for stars of a given surface temperature.  A dependence on metallicity was also found, particularly for the 1.19\mic~and 1.28\mic~blends, with stars of higher (lower) metallicity having elevated (diminished) equivalent widths.   

The classification of a star, even in the presence of strong reddening, is possible by obtaining the equivalent widths of four blends, or the equivalent widths of two blends and the slope of the stellar spectrum.  The ratio of Mg I (or P$\gamma$, though it is less reliable) to 1.16\mic~indicates the surface gravity of the star, with a higher ratio for lower surface gravity stars.  The temperature can be estimated using the equivalent width of the 1.28\mic~blend or the CN index, or alternatively the slope of the spectrum can be used.  Measuring these temperature- and gravity-sensitive features in integrated spectra of composite stellar populations (in star clusters or galaxies for example) can therefore provide constraints on the age and other astrophysical properties of the system. 

\acknowledgements

We thank Don Figer for help with an IRAF script for spectroscopic data reduction.

\references

\reference{} Cayrel de Strobel, G. Soubiran, C., Friel, E. D., Ralite, N., Francois, P. 1997, \aaps, 124, 299
\reference{} Fanson, J. L. 1998, SPIE, 3356, 478
\reference{} Figer, Donald F., McLean, Ian S., Morris, Mark 1999, \apj, 514, 202
\reference{} Hicks, E. K., Malkan, M. A., Teplitz, H. I., Sugai, H.,
 Guichard, J. 2000, AAS, 197, 4408
\reference{} Hinkle, K., Wallace, L., Livingston, W., 1995, \pasp, 107, 1042
\reference{} Jones, H. R. A., Longmore, A. J., Allard, F., and Hauschildt, P. H. 1996, \mnras, 280, 77
\reference{} Lancon, A. and Rocca-Volmerange, B. 1992, \aaps, 96, 593
\reference{} McLean, Ian S., Macintosh, Bruce A., Liu, Tim, Casement, L. S., Figer, Donald F., Lacayanga, Fred, Larson, Sam, Teplitz, Harry, Silverstone, Murray, Becklin, Eric E. 1994 SPIE, 2198, 457
\reference{} McLean, I. S., 1994, in {\it Infrared Astronomy with Arrays: the Next Generation}, ed. McLean, I.S. (Kluwer)
\reference{} McLean, I. S., et al. 1998, Proc. SPIE, 3354, 566
\reference{} McLean, I. S., \et 2000, \apjl, 533, L45
\reference{} Mouhcine, M., Lancon, A. 2001 ApSSS, 277, 485
\reference{} Nesterov, V. V., \et 1995, \apjs, 110, 367
\reference{} Stockman, H.S., 1997, ed., ``Next Generation Space
Telescope: Visiting a Time When Galaxies Were Young'', (AURA, Inc.:
Baltimore)
\reference{} Sugai, H., Malkan, M. A., Ward, M. J., Davies, R. I., McLean, I. S. 1997, \apj, 481, 186
\reference{} Wallace, L., Livingston, W., Hinkle, K., Bernath, P., 1996, \apjs, 106, 165
\reference{} Wallace, Lloyd, Meyer, Michael R., Hinkle, Kenneth, Edwards, Suzan 2000, \apj, 535, 325
\reference{} Wallace, Lloyd and Hinkle, Kenneth 1996, \apjs, 107, 312

\clearpage

\clearpage														
\begin{deluxetable}{rlllll|rllll}														
\tabletypesize{\footnotesize}														
\tablenum{1}														
\tablecolumns{11}														
\tablewidth{0pc}														
\tablecaption{J-Band Atlas Stars}														
\tablehead{														
\cline{1-11}\\														
\colhead{HR \#} & \colhead{Type}& \colhead{Teff(K)} &\colhead{Date Obs.} & \colhead{[Fe/H]\tablenotemark{a}}& \colhead{} & \colhead{HR \#} & \colhead{Type} & \colhead{Teff(K)} &\colhead{Date Obs.} & \colhead{[Fe/H]\tablenotemark{a}}}														
\startdata

81	&A0V	&9480	&	1999 Sep 19	&\nodata	&	&	4886	&A7V	&7930	&	1996 Mar 14	&\nodata 	\\
106	&K5III	&3980	&	1996 Aug 26	&\nodata	&	&	4900	&A7III	&7650	&	1996 Mar 14	&\nodata 	\\
207	&G0Ib	&5510	&	1999 Sep 19	&\nodata	&	&	4943	&B9V	&10700	&	1996 Mar 14	&\nodata 	\\
217	&F8V	&6135	&	1996 Aug 26	&\nodata	&	&	4967	&B7III	&13200	&	1996 Mar 14	&\nodata 	\\
225	&F8V	&6135	&	1994 Dec 16	&\nodata	&	&	5191	&B3V	&19000	&	1999 Sep 18	&\nodata 	\\
232	&A3V	&8595	&	1996 Aug 26	&\nodata	&	&	5322	&F9V	&6035	&	1996 Mar 13	&0.20	\\
246	&A2V	&8810	&	1996 Aug 26	&\nodata	&	&	5747	&F0Vp	&7020	&	1999 Sep 17	&0.825	\\
601	&M2III	&3710	&	1999 Sep 19	&\nodata	&	&	5770	&B9V	&10700	&	1997 Mar 30	&\nodata 	\\
611	&K5Iab	&3850	&	1996 Aug 26	&\nodata	&	&	6088	&A3V	&8595	&	1997 Oct 13	&\nodata 	\\
617	&K2III	&4260	&	1999 Sep 18	&-0.258	&	&	6156	&A1V	&9150	&	1996 Aug 25	&\nodata 	\\
690	&F7Ib	&6370	&	1999 Sep 19	&\nodata	&	&	6466	&G0III	&5910	&	1999 Sep 18	&\nodata 	\\
695	&G2V	&5830	&	1996 Aug 26	&-0.10	&	&	6518	&K0V	&5240 	&	1999 Sep 18	&\nodata	\\
719	&K0III	&4810	&	1999 Sep 17	&\nodata	&	&	6598	&F9V	&6035 	&	1997 Oct 13	&-0.39  	\\
747	&K5Iab	&3850	&	1999 Sep 18	&\nodata	&	&	6604	&F5II	&6640 	&	1999 Sep 17	& 0.55  	\\
753	&K3V	&4790	&	1999 Sep 18	&\nodata	&	&	6685	&F2Ibe	&7170 	&	1999 Sep 17	&-0.41  	\\
867	&M6III	&3250	&	1999 Sep 19	&\nodata	&	&	6985	&F5III	&6470 	&	1999 Sep 18	&0.10   	\\
921	&M4II	&2980	&	1999 Sep 19	&0.05	&	&	7139	&M4III	&3560 	&	1999 Sep 17	&\nodata	\\
940	&M0III	&3820	&	1999 Sep 18	&\nodata	&	&	7345	&G8V	&5430 	&	1999 Sep 18	&\nodata	\\
1542	&O9.5I	&3250	&	1994 Dec 17	0&0.30	&	&	7373	&G8IV	&5430 	&	1999 Sep 18	&0.38   	\\
1544	&A1V	&9150	&	1997 Oct 13	&\nodata	&	&	7384	&A0V	&9480 	&	1999 Sep 17	&\nodata	\\
1562	&M1III	&3780	&	1999 Sep 18	&\nodata	&	&	7471	&B3III	&17100	&	1999 Sep 19	&\nodata	\\
1683	&A0V	&9480	&	1995 Jan 19	&\nodata	&	&	7475	&K4Ib	&3990 	&	1999 Sep 17	&\nodata	\\
1684	&K5III	&3980	&	1997 Oct 13	&-0.11	&	&	7563	&F0III	&7150 	&	1999 Sep 18	&\nodata	\\
1703	&M0V	&3800	&	1999 Sep 19	&\nodata	&	&	7589	&O9.5I	&32500	&	1996 Mar 14	&\nodata	\\
1988	&G4V	&5740	&	1997 Oct 13	&\nodata	&	&	7596	&A0III	&10100	&	1999 Sep 18	&\nodata	\\
2007	&G4V	&5740	&	1997 Oct 15	&\nodata	&	&	7696	&M3III	&3630 	&	1999 Sep 18	&\nodata	\\
2034	&A0V	&9480	&	1999 Sep 19	&\nodata	&	&	7776	&K0:II:+AII	&4420 	&	1999 Sep 17	&0.62   \\
2208	&G2V	&5830	&	1997 Oct 14	&\nodata	&	&	7800	&K7III	&3920	&	1999 Sep 18	&\nodata	\\
2285	&A4V	&8375	&	1995 Jan 19	&\nodata	&	&	8143	&B9Iab	&10500	&	1997 Oct 12	&\nodata	\\
2375	&A3V	&8595	&	1997 Oct 14	&\nodata	&	&	8149	&K5III	&3980 	&	1994 Dec 16	&\nodata	\\
2401	&F8V	&6135	&	1995 Dec 9	&\nodata	&	&	8279	&B2Ib	&17800	&	1997 Mar 30	&-0.33  	\\
3257	&A2V	&8810	&	1994 Dec 17	&\nodata	&	&	8328	&A1V	&9150 	&	1997 Oct 12	&\nodata	\\
3436	&G8III	&4960	&	1995 Jan 19	&\nodata 	&	&	8334	&A2Iab	&9080	&	1999 Sep 18	&\nodata 	\\
3592	&A3V	&8595	&	1997 Mar 29	&\nodata 	&	&	8345	&A2Ib	&9080	&	1999 Sep 18	&\nodata 	\\
3608	&A2V	&8810	&	1994 Dec 16	&\nodata 	&	&	8518	&K4V	&4560	&	1996 Mar 13	&0.3     	\\
3719	&A5V	&8160	&	1996 Mar 9	&\nodata 	&	&	8631	&G4V	&5740	&	1994 Dec 15	&\nodata 	\\
3750	&G2V	&5830	&	1994 Dec 17	&-0.31	&	&	8692	&G4Ib	&4900	&	1999 Sep 19	&\nodata 	\\
3881	&G0.5Va	&5860	&	1995 Jan 18	&0.0	&	&	8718	&F5II	&6640	&	1999 Sep 19	&0.045   	\\
3998	&F7V	&6390	&	1995 Jan 18	&0.0	&	&	8726	&K5Ib	&3850	&	1999 Sep 19	&0.015   	\\
4012	&F9V	&6035	&	1995 Jan 20	&0.115	&	&	8738	&A1V	&9150	&	1994 Dec 15	&\nodata 	\\
4026	&A8III	&7320	&	1994 Dec 17	&\nodata 	&	&	8826	&A5Vn	&8160	&	1997 Oct 13	&\nodata 	\\
4051	&F9V	&6035	&	1996 Mar 9	&\nodata 	&	&	8852	&G9III:Fe2III	&4890	&	1999 Sep 19	&-0.382  \\
4215	&A1V	&9150	&	1995 Jan 19	&\nodata 	&	&	9010	&K3II	&4130	&	1999 Sep 19	&\nodata 	\\
4412	&F7V	&6240	&	1996 Mar 9	&\nodata 	&	&	9053	&G8Ib	&4590	&	1999 Sep 19	&\nodata 	\\
4439	&F8V	&6135	&	1994 Dec 16	&\nodata 	&	&	9066	&M7IIIe	&3080	&	1999 Sep 19	&\nodata 	\\
4454	&A5V	&8160	&	1995 Jan 18	&\nodata 	&	&	9099	&M4III	&3560	&	1999 Sep 19	&\nodata 	\\
4632	&A4V	&8375	&	1996 Mar 9	&\nodata 	&	&	 HD10986 	&K5III	&3980	&	1996 Aug 26	&\nodata \\
4660	&A3V	&8595	&	1994 Dec 17	&\nodata 	&	&	 HD11469	&M3.5I	&3090	&	1996 Aug 26	&\nodata \\
4663	&A3V	&8595	&	1997 Mar 30	&\nodata 	&	&	 HD18881	&A0V	&9480	&	1996 Oct 26	&\nodata \\
4708	&F8V	&6135	&	1995 Jan 18	&\nodata 	&	&	 HD36395 	&M1.5V	&3600	&	1999 Sep 18	&0.60    \\
4767	&F9V	&6035	&	1994 Dec 17	&-0.11	&	&	 HD162208	&A0V	&9480	&	1996 Aug 25	&\nodata 	\\
4861	&A1V	&9150	&	1996 Mar 14	&\nodata 	&	&	 HD184314	&K5V	&4340	&	1996 Aug 25	&\nodata \\
4875	&A3V	&8595	&	1996 Mar 14	&\nodata 	&	&	 	&                     	&	&		&         \\
														
\enddata														
\tablenotetext{a}{Cayrel de Strobel, G., et al. 1997, A\&AS, 124, 299}														
\end{deluxetable}														

\clearpage							
\begin{deluxetable}{lllllll} 							
\tabletypesize{\footnotesize}
\rotate							
\tablenum{2}							
\tablecolumns{6}  							
\tablewidth{0pc}  							
\tablecaption{Spectral Features Identified}  							
\tablehead{  							
\colhead{} & \multicolumn{2}{c}{Dominant Features} & \colhead{} & \colhead{} & \colhead{} \\							
\cline{2-3} \\							
\colhead{Name} & \colhead{Element(s)} & \colhead{$\lambda$(\mic)} & \colhead{Blend $\lambda$ (K-M $\lambda$)\tablenotemark{a}} & \colhead{Comments} & \colhead{Cool Stars}}							
\startdata

P $\gamma$	  &	H I	&	&1.0942	&	&Not present in K and M stars	\\
CN 0-0 index	&	CN	&	&1.0960-1.1040	&	&Present in T$_{eff} \le$  5500 K	\\
1.16	  &	Si I/Si I/Fe I 	&1.1595	&1.1600	&	&Fe strongest, CN 0-0	\\
	  &	Fe I/Cr I/Si I 	&1.1612	&	&	&present	\\
1.17	  &	C I (3 lines)   	&1.1753	&1.1775	&	&C weaker, CN 0-0 present	\\
	  &	K I/K I/C I/Fe I	&1.1780	&	&	&K I in late M	\\
Mg I	  &	Mg I	&	&1.1831	&	&	\\
1.19	  &	Fe I 	&1.1886	&1.1886 (1.1875)	&Several C I lines	&C weaker, Ti I(1.1896)	\\
	  &	Fe I	&1.1887	&	&	&present, CN 0-0 present	\\
1.20	  &	Fe I	&1.1976	&1.1985 (1.1980)	&Si I(1.1987) strongest	&Ti I(1.1977) present	\\
	  &	Si I	&1.1987	&	&	&FeH trough in late M	\\
	  &	Si I	&1.1995	&	&	&	\\
Si I	  &	Si I	&	&1.2034	&	&	\\
1.21	  &	Si I	&1.2086	&1.2100 (1.2090)	&Si I(1.2086) and  	&Si I(1.2106) weaker	\\
	  &	Mg I	&1.2087	&	&Mg I(1.2106) strongest	&	\\
	  &	Si I	&1.2106	&	&	&	\\
	  &	Si I	&1.2114	&	&	&	\\
1.28	  &	H I	&	&1.2822 (1.2830)	&	&Ti I(1.2825) present	\\
1.31	  &	Si I	&1.3105	&1.3145	&Al I(1.3126) and	&	\\
	  &	Al I	&1.3126	&	&Si I(1.3180) strongest	&	\\
	  &	Al I	&1.3154	&	&	&	\\
	  &	Si I	&1.3180	&	&	&	\\
1.33	  &	Si I	&1.3291	&1.3320 (1.3310)	&Si I(1.3291),Si I(1.3313)	&Fe and Ca lines weaker	\\
	  &	Si I	&1.3313	&	&and Fe I/Si I(1.3330) 	&and Mn lines strengther	\\
	  &	Fe I/Si I	&1.3330	&	&strongest	&H$_{2}$O absn in late M	\\
							
\enddata							
\tablenotetext{a}{If a shift in central wavelength with change in temperature is present, then the central wavelength for cooler stars (K and M stars) is given in parenthesis.}							
\end{deluxetable}							

\clearpage																											
\begin{deluxetable}{llllllllllllll} 																											
\tabletypesize{\footnotesize}																									
\tablenum{3}																											
\tablecolumns{14}  																											
\tablewidth{0pc}  																											
\tablecaption{Equivalent Widths of Each Line/Blend}  																											
\tablehead{  																											
\colhead{HR  \#} & \colhead{ST} & \colhead{P$\gamma$} & \colhead{CN 0-0} & \colhead{1.16} & \colhead{1.17} & \colhead{Mg I} & \colhead{1.19} & \colhead{1.20} & \colhead{Si I} & \colhead{1.21} & \colhead{1.28} & \colhead{1.31} & \colhead{1.33}\tablenotemark{c}}																											
\startdata																											
																											
1542	&	O9.5I	&	2.61\tablenotemark{b}	&	\nodata	&	\nodata	&	\nodata	&	\nodata	&	\nodata	&	\nodata	&	\nodata	&	\nodata	&	0.61	&	\nodata	&	\nodata	\\
7589	&	O9.5I	&	13.91	&	\nodata	&	\nodata	&	\nodata	&	\nodata	&	\nodata	&	\nodata	&	\nodata	&	\nodata	&	1.84	&	\nodata	&	\nodata	\\
8279	&	B2Ib	&	2.67	&	\nodata	&	\nodata	&	\nodata	&	\nodata	&	\nodata	&	\nodata	&	\nodata	&	\nodata	&	7.39:\tablenotemark{b}	&	\nodata	&	\nodata	\\
5191	&	B3	&	6.24	&	\nodata	&	\nodata	&	\nodata	&	\nodata	&	\nodata	&	\nodata	&	\nodata	&	\nodata	&	2.88	&	\nodata	&	\nodata	\\
7471	&	B3III	&	9.08:\tablenotemark{b}	&	\nodata	&	\nodata	&	\nodata	&	\nodata	&	\nodata	&	\nodata	&	\nodata	&	\nodata	&	7.46:\tablenotemark{b}	&	\nodata	&	\nodata	\\
4967	&	B7III	&	6.16	&	\nodata	&	\nodata	&	\nodata	&	\nodata	&	\nodata	&	\nodata	&	\nodata	&	\nodata	&	5.79	&	\nodata	&	\nodata	\\
4943	&	B9V	&	8.77	&	\nodata	&	\nodata	&	\nodata	&	\nodata	&	\nodata	&	\nodata	&	\nodata	&	\nodata	&	7.44:	&	\nodata	&	\nodata	\\
5770	&	B9V	&	21.62:	&	\nodata	&	\nodata	&	\nodata	&	\nodata	&	\nodata	&	\nodata	&	\nodata	&	\nodata	&	8.06:	&	\nodata	&	\nodata	\\
8143	&	B9Iab	&	3.09	&	\nodata	&	\nodata	&	\nodata	&	\nodata	&	\nodata	&	\nodata	&	\nodata	&	\nodata	&	4.27\tablenotemark{b}	&	\nodata	&	\nodata	\\
1683	&	A0V	&	7.09	&	\nodata	&	\nodata	&	\nodata	&	\nodata	&	\nodata	&	\nodata	&	\nodata	&	\nodata	&	8.34:	&	\nodata	&	\nodata	\\
81	&	A0V	&	7.75	&	\nodata	&	\nodata	&	\nodata	&	\nodata	&	\nodata	&	\nodata	&	\nodata	&	\nodata	&	7.13	&	\nodata	&	\nodata	\\
HD18881	&	A0V	&	7.96	&	\nodata	&	\nodata	&	\nodata	&	\nodata	&	\nodata	&	\nodata	&	\nodata	&	\nodata	&	7.54	&	\nodata	&	\nodata	\\
HD162208	&	A0V	&	7.05	&	\nodata	&	\nodata	&	\nodata	&	\nodata	&	\nodata	&	\nodata	&	\nodata	&	\nodata	&	7.68	&	\nodata	&	\nodata	\\
7384	&	A0V	&	9.76:	&	\nodata	&	\nodata	&	\nodata	&	\nodata	&	\nodata	&	\nodata	&	\nodata	&	\nodata	&	13.02:	&	\nodata	&	\nodata	\\
2034	&	A0V	&	7.55:	&	\nodata	&	\nodata	&	\nodata	&	\nodata	&	\nodata	&	\nodata	&	\nodata	&	\nodata	&	11.89:\tablenotemark{b}	&	\nodata	&	\nodata	\\
7596	&	A0III	&	10.49:	&	\nodata	&	\nodata	&	\nodata	&	\nodata	&	\nodata	&	\nodata	&	\nodata	&	\nodata	&	10.62:\tablenotemark{b}	&	\nodata	&	\nodata	\\
4215	&	A1V	&	7.18	&	\nodata	&	\nodata	&	\nodata	&	\nodata	&	\nodata	&	\nodata	&	\nodata	&	\nodata	&	7.44:	&	\nodata	&	\nodata	\\
4861	&	A1V	&	6.27	&	\nodata	&	\nodata	&	\nodata	&	\nodata	&	\nodata	&	\nodata	&	\nodata	&	\nodata	&	7.07	&	\nodata	&	\nodata	\\
8738	&	A1V	&	7.27	&	\nodata	&	\nodata	&	\nodata	&	\nodata	&	\nodata	&	\nodata	&	\nodata	&	\nodata	&	8.63	&	\nodata	&	\nodata	\\
6156	&	A1V	&	2.10\tablenotemark{a}	&	\nodata	&	\nodata	&	\nodata	&	\nodata	&	\nodata	&	\nodata	&	\nodata	&	\nodata	&	7.08:	&	\nodata	&	\nodata	\\
8328	&	A1V	&	8.78:\tablenotemark{b}	&	\nodata	&	\nodata	&	\nodata	&	\nodata	&	\nodata	&	\nodata	&	\nodata	&	\nodata	&	6.04:	&	\nodata	&	\nodata	\\
1544	&	A1V	&	7.86	&	\nodata	&	\nodata	&	\nodata	&	\nodata	&	\nodata	&	\nodata	&	\nodata	&	\nodata	&	6.04:	&	\nodata	&	\nodata	\\
3608	&	A2V	&	9.46	&	\nodata	&	\nodata	&	\nodata	&	\nodata	&	\nodata	&	\nodata	&	\nodata	&	\nodata	&	6.41	&	\nodata	&	\nodata	\\
3257	&	A2V	&	6.40	&	\nodata	&	\nodata	&	\nodata	&	\nodata	&	\nodata	&	\nodata	&	\nodata	&	\nodata	&	6.56	&	\nodata	&	\nodata	\\
246	&	A2V	&	4.97	&	\nodata	&	\nodata	&	\nodata	&	\nodata	&	\nodata	&	\nodata	&	\nodata	&	\nodata	&	4.50	&	\nodata	&	\nodata	\\
8334	&	A2Iab	&	2.30	&	\nodata	&	\nodata	&	\nodata	&	\nodata	&	\nodata	&	\nodata	&	\nodata	&	\nodata	&	4.45\tablenotemark{b}	&	\nodata	&	\nodata	\\
8345	&	A2Ib	&	2.23\tablenotemark{b}	&	\nodata	&	\nodata	&	\nodata	&	\nodata	&	\nodata	&	\nodata	&	\nodata	&	\nodata	&	3.94	&	\nodata	&	\nodata	\\
4663	&	A3V	&	13.11:	&	\nodata	&	\nodata	&	\nodata	&	\nodata	&	\nodata	&	\nodata	&	\nodata	&	\nodata	&	9.02:	&	\nodata	&	\nodata	\\
4875	&	A3V	&	6.87	&	\nodata	&	\nodata	&	\nodata	&	\nodata	&	\nodata	&	\nodata	&	\nodata	&	\nodata	&	6.55	&	\nodata	&	\nodata	\\
4660	&	A3V	&	6.51	&	\nodata	&	\nodata	&	\nodata	&	\nodata	&	\nodata	&	\nodata	&	\nodata	&	\nodata	&	5.00	&	\nodata	&	\nodata	\\
6088	&	A3V	&	5.86	&	\nodata	&	\nodata	&	\nodata	&	\nodata	&	\nodata	&	\nodata	&	\nodata	&	\nodata	&	7.34:	&	\nodata	&	\nodata	\\
2375	&	A3V	&	7.55	&	\nodata	&	\nodata	&	\nodata	&	\nodata	&	\nodata	&	\nodata	&	\nodata	&	\nodata	&	7.41:	&	\nodata	&	\nodata	\\
232	&	A3V	&	5.95\tablenotemark{b}	&	\nodata	&	\nodata	&	\nodata	&	\nodata	&	\nodata	&	\nodata	&	\nodata	&	\nodata	&	6.75	&	\nodata	&	\nodata	\\
3592	&	A3V	&	6.35	&	\nodata	&	\nodata	&	\nodata	&	\nodata	&	\nodata	&	\nodata	&	\nodata	&	\nodata	&	9.87:	&	\nodata	&	\nodata	\\
2285	&	A4V	&	5.76	&	\nodata	&	\nodata	&	\nodata	&	\nodata	&	\nodata	&	\nodata	&	\nodata	&	\nodata	&	7.04:	&	\nodata	&	\nodata	\\
4632	&	A4V	&	5.63	&	\nodata	&	1.42	&	1.12	&	2.17	&	1.63	&	1.23	&	1.18	&	1.14:	&	7.26	&	1.38	&	2.43	\\
4454	&	A5V	&	5.09	&	\nodata	&	\nodata	&	\nodata	&	\nodata	&	\nodata	&	\nodata	&	\nodata	&	\nodata	&	6.17	&	\nodata	&	\nodata	\\
3719	&	A5V	&	5.28	&	\nodata	&	1.51	&	1.03	&	3.79:	&	1.33	&	1.41\tablenotemark{a}	&	1.37	&	1.41\tablenotemark{a}	&	7.76	&	1.01	&	2.30	\\
8826	&	A5Vn	&	6.17	&	\nodata	&	\nodata	&	\nodata	&	\nodata	&	\nodata	&	\nodata	&	\nodata	&	\nodata	&	6.00	&	\nodata	&	\nodata	\\
4886	&	A7V	&	4.18	&	\nodata	&	1.03	&	1.41\tablenotemark{a}	&	1.34:	&	1.50:	&	0.79:	&	1.13	&	0.53	&	5.72	&	1.08	&	2.02	\\
4900	&	A7III	&	4.72	&	\nodata	&	1.74	&	1.18	&	1.41\tablenotemark{a}	&	1.41\tablenotemark{a}	&	1.63	&	1.93	&	1.16	&	5.28	&	0.56	&	1.30:	\\
4026	&	A8III	&	7.45	&	\nodata	&	1.79:	&	0.91	&	2.85	&	1.89	&	1.59	&	1.75	&	1.02:	&	7.67	&	1.01	&	5.43:	\\
5747	&	F0Vp	&	5.35	&	\nodata	&	6.40:	&	1.09	&	3.65	&	3.22	&	1.59	&	2.41	&	0.60	&	14.73:	&	4.17	&	8.14	\\
7563	&	F0III	&	2.87	&	\nodata	&	2.61	&	1.57	&	2.67	&	1.97	&	1.51	&	1.23	&	1.41\tablenotemark{a}	&	7.86	&	2.63	&	3.74	\\
6685	&	F2Ibe	&	2.20:\tablenotemark{b}	&	\nodata	&	1.37	&	3.11\tablenotemark{b}	&	1.03	&	1.18	&	0.58	&	0.96	&	0.97	&	4.41:	&	1.90	&	3.48	\\
6604	&	F5II	&	3.23	&	\nodata	&	1.08	&	4.14:\tablenotemark{b}	&	2.93	&	4.30	&	1.70	&	2.26	&	1.16	&	11.68:	&	4.05	&	6.06	\\
8718	&	F5II	&	7.19:	&	\nodata	&	4.35	&	1.11	&	0.85	&	1.75	&	0.98	&	1.48\tablenotemark{b}	&	0.31	&	7.45:\tablenotemark{b}	&	1.32\tablenotemark{b}	&	1.16:	\\
6985	&	F5III	&	2.96	&	\nodata	&	3.31:\tablenotemark{b}	&	0.84	&	1.57	&	5.44:	&	1.28	&	0.81	&	1.10	&	7.70:\tablenotemark{b}	&	1.86\tablenotemark{b}	&	3.10	\\
3998	&	F7V	&	1.77	&	\nodata	&	1.91	&	1.01	&	1.65:	&	0.90	&	1.12	&	0.92	&	1.61:	&	3.88	&	1.35	&	2.61	\\
4412	&	F7V	&	1.87	&	\nodata	&	1.77	&	0.84	&	1.49	&	1.07	&	1.07	&	1.05	&	1.45	&	4.30	&	1.29	&	3.29	\\
690	&	F7Ib	&	4.64:\tablenotemark{b}	&	\nodata	&	2.09:	&	2.17	&	1.04	&	1.00	&	1.57	&	0.84	&	1.41\tablenotemark{a}	&	4.97\tablenotemark{b}	&	1.30\tablenotemark{b}	&	0.84:	\\
4439	&	F8V	&	1.71	&	\nodata	&	2.03	&	0.86	&	3.66:	&	1.75	&	1.02	&	1.25	&	0.81	&	2.42	&	1.56	&	4.87	\\
225	&	F8V	&	1.45	&	\nodata	&	1.31	&	1.41\tablenotemark{a}	&	1.53	&	1.69	&	1.40	&	0.68	&	1.41\tablenotemark{a}	&	2.48	&	0.81	&	2.94	\\
4708	&	F8V	&	1.79	&	\nodata	&	1.76	&	0.95	&	1.89	&	1.30	&	1.05	&	1.38	&	1.61	&	3.29	&	1.08:	&	2.73	\\
2401	&	F8V	&	0.97	&	\nodata	&	2.29:	&	1.09	&	1.83	&	1.23	&	0.74	&	1.63	&	1.87:	&	2.94	&	0.65	&	1.32	\\
217	&	F8V	&	2.44:\tablenotemark{b}	&	\nodata	&	1.62:	&	0.97	&	1.06	&	1.10	&	0.88	&	1.01	&	0.80	&	3.19	&	1.47	&	1.63	\\
4012	&	F9V	&	1.42	&	\nodata	&	1.28	&	1.25	&	1.03	&	0.84	&	0.64	&	0.54	&	0.94	&	3.20	&	1.07:	&	0.93	\\
5322	&	F9V	&	2.52:	&	\nodata	&	1.39:	&	1.41\tablenotemark{a}	&	1.85	&	2.03	&	3.53:	&	1.71:	&	1.61:	&	2.99	&	1.61	&	3.04	\\
4051	&	F9V	&	2.38	&	\nodata	&	2.76	&	1.27	&	1.41\tablenotemark{a}	&	1.42	&	0.76	&	1.28	&	1.24	&	3.33	&	1.58	&	2.79	\\
4767	&	F9V	&	1.80	&	\nodata	&	1.68:	&	0.98	&	2.41	&	1.51	&	1.09	&	1.44	&	0.98	&	2.96	&	1.16	&	3.43	\\
6598	&	F9V	&	1.12	&	\nodata	&	0.47:	&	0.29	&	1.12	&	0.31:	&	1.25:	&	0.92:	&	0.59:\tablenotemark{b}	&	1.87	&	0.36:	&	0.92	\\
207	&	G0Ib	&	6.41:\tablenotemark{b}	&	-0.42	&	3.08\tablenotemark{b}	&	1.26	&	0.85	&	0.52	&	1.71\tablenotemark{b}	&	0.97\tablenotemark{b}	&	1.41\tablenotemark{a}	&	4.16:\tablenotemark{b}	&	1.77\tablenotemark{b}	&	3.04	\\
6466	&	G0III	&	9.69:\tablenotemark{b}	&	-0.95	&	2.33	&	1.00	&	2.19\tablenotemark{b}	&	1.93	&	1.30	&	0.79	&	1.41\tablenotemark{a}	&	4.00:\tablenotemark{b}	&	1.10\tablenotemark{b}	&	1.40:	\\
3881	&	G0.5Va	&	0.84	&	\nodata	&	1.77	&	0.40	&	1.17:	&	1.19	&	1.08	&	1.48	&	1.60:	&	2.29	&	1.31	&	2.39	\\
695	&	G2V	&	0.52:	&	-0.03	&	1.27	&	0.23:	&	1.68:	&	1.09	&	1.41	&	1.41\tablenotemark{a}	&	0.45:	&	2.45	&	0.86:	&	2.75	\\
3750	&	G2V	&	1.39:	&	3.92	&	1.46	&	2.46:	&	3.45:	&	1.97	&	0.98	&	1.61	&	0.91	&	1.07	&	0.95	&	4.59	\\
2208	&	G2V	&	1.12	&	0.21	&	1.99	&	0.87:	&	2.28	&	1.33	&	2.68:	&	1.53	&	1.72	&	2.29	&	1.41\tablenotemark{a}	&	\nodata	\\
8631	&	G4V	&	0.85:	&	0.62	&	1.77	&	2.47:	&	2.58	&	1.57	&	1.46	&	1.09	&	2.46:	&	1.15	&	1.12	&	5.37:	\\
1988	&	G4V	&	1.74	&	\nodata	&	1.45	&	1.03:	&	0.89	&	1.35	&	1.05	&	0.88	&	1.13:	&	1.92	&	0.29:	&	1.08	\\
2007	&	G4V	&	1.95:	&	-3.18	&	0.63:\tablenotemark{b}	&	0.83	&	1.54	&	0.88:	&	0.90	&	0.60	&	0.69:	&	2.06	&	0.51	&	3.28	\\
8692	&	G4Ib	&	16.80:\tablenotemark{b}	&	3.66	&	4.77:\tablenotemark{b}	&	1.26	&	1.27\tablenotemark{b}	&	2.28	&	3.10	&	2.36\tablenotemark{b}	&	1.36	&	2.77:	&	2.16\tablenotemark{b}	&	4.58	\\
7345	&	G8V	&	4.45	&	-1.56	&	2.60:\tablenotemark{b}	&	1.10	&	1.84\tablenotemark{b}	&	2.43	&	2.32	&	1.41\tablenotemark{b}	&	1.58	&	4.10\tablenotemark{b}	&	2.51\tablenotemark{b}	&	2.80	\\
9053	&	G8Ib	&	17.63:\tablenotemark{b}	&	\nodata	&	3.66:\tablenotemark{b}	&	1.29	&	1.11\tablenotemark{b}	&	1.98	&	2.18	&	0.93\tablenotemark{b}	&	1.41	&	3.57:	&	1.96\tablenotemark{b}	&	3.38	\\
3436	&	G8III	&	1.65:	&	\nodata	&	1.41\tablenotemark{a}	&	0.91	&	1.12	&	0.61	&	0.98	&	0.98\tablenotemark{b}	&	0.98	&	1.43	&	0.49:\tablenotemark{b}	&	\nodata	\\
7373	&	G8IV	&	8.13:\tablenotemark{b}	&	-0.62	&	3.33:	&	2.63:\tablenotemark{b}	&	2.05	&	2.84	&	1.86	&	2.11	&	1.22	&	4.94:	&	2.93	&	5.50	\\
8852	&	G9III:Fe2	&	3.63\tablenotemark{b}	&	-0.07	&	4.17	&	2.00	&	3.08	&	3.37	&	2.29	&	1.42	&	1.82:	&	4.27:\tablenotemark{b}	&	3.06	&	6.13	\\
6518	&	K0V	&	\nodata	&	2.17	&	2.01	&	0.79	&	2.32\tablenotemark{b}	&	0.89	&	1.98	&	1.76\tablenotemark{b}	&	1.15	&	3.08:\tablenotemark{b}	&	2.31\tablenotemark{b}	&	3.38	\\
7776	&	K0:II:+A5:N 	&	\nodata	&	54.99:	&	9.01:	&	2.04	&	3.49	&	7.85:\tablenotemark{b}	&	1.84	&	2.84	&	0.74	&	3.62	&	5.83	&	15.22	\\
719	&	K0III	&	\nodata	&	0.95	&	4.51	&	0.92	&	3.10	&	4.43	&	1.56	&	2.26	&	0.99	&	4.63:	&	4.17	&	12.24	\\
617	&	K2III	&	\nodata	&	1.67	&	4.38	&	0.94	&	3.54	&	0.95	&	1.75	&	2.42	&	0.51	&	2.48	&	5.17	&	8.78	\\
753	&	K3V	&	\nodata	&	5.92:	&	3.19	&	1.11	&	2.71	&	3.43:	&	1.44	&	1.44	&	1.41\tablenotemark{a}	&	3.66:	&	4.31	&	6.26	\\
9010	&	K3IIb	&	\nodata	&	6.94	&	5.50:	&	1.55	&	3.10\tablenotemark{b}	&	1.32\tablenotemark{b}	&	1.25	&	1.09\tablenotemark{b}	&	0.85	&	3.54:	&	2.06\tablenotemark{b}	&	3.49	\\
8518	&	K4V	&	\nodata	&	2.54	&	1.98	&	1.50	&	3.14:	&	1.60	&	1.25:	&	1.21	&	1.40	&	0.61:	&	2.62:	&	4.73	\\
7475	&	K4Ib	&	\nodata	&	7.48	&	6.22	&	1.66	&	3.90	&	4.49	&	2.81	&	3.54	&	1.41\tablenotemark{a}	&	5.25:	&	4.76	&	8.36	\\
HD184314 	&	K5V	&	\nodata	&	1.61	&	2.90	&	1.45	&	2.50	&	2.21	&	1.41	&	1.73	&	1.92:	&	1.46	&	0.96	&	1.64	\\
 HD10986 	&	K5III	&	\nodata	&	6.19	&	1.90	&	1.42	&	2.36	&	2.32	&	1.39	&	0.74	&	2.04	&	1.00	&	1.97	&	1.41	\\
106	&	K5III	&	\nodata	&	3.96	&	1.84	&	1.13:	&	1.41\tablenotemark{a}	&	1.22	&	1.22:	&	0.67	&	1.41\tablenotemark{a}	&	1.37	&	1.41:	&	2.00	\\
1684	&	K5III	&	\nodata	&	3.81	&	2.41:	&	1.07	&	1.22	&	1.28	&	1.37	&	0.28:	&	0.72	&	1.33	&	0.99	&	2.07	\\
8149	&	K5III	&	\nodata	&	1.69	&	1.87	&	2.61:	&	1.93	&	1.87	&	0.79	&	0.55	&	0.84	&	1.68	&	0.78	&	1.76	\\
747	&	K5Iab	&	\nodata	&	4.24	&	1.41\tablenotemark{a}	&	1.55	&	1.07	&	1.41\tablenotemark{a}	&	1.09:	&	0.81	&	1.69:	&	1.18	&	1.32	&	1.30	\\
611	&	K5Iab	&	\nodata	&	3.21	&	2.30	&	1.40	&	1.84	&	2.34	&	1.40	&	1.19	&	2.18	&	4.00	&	1.44	&	2.10	\\
8726	&	K5Ib	&	\nodata	&	4.82	&	7.50	&	1.80	&	1.35	&	2.22	&	4.15:	&	1.11	&	1.60:\tablenotemark{b}	&	5.53:\tablenotemark{b}	&	4.09	&	4.53	\\
7800	&	K7III	&	\nodata	&	2.00	&	2.56:	&	3.78:	&	0.33	&	0.65	&	0.60	&	0.78	&	1.41\tablenotemark{a}	&	3.49:\tablenotemark{b}	&	1.44	&	3.13	\\
1703	&	M0V	&	\nodata	&	4.50	&	3.37:	&	2.56:	&	0.36	&	0.35	&	0.50	&	0.58	&	1.41\tablenotemark{a}	&	3.62:\tablenotemark{b}	&	1.20	&	3.05	\\
940	&	M0III	&	\nodata	&	1.13	&	5.27	&	1.90	&	2.73	&	2.75	&	1.21	&	1.13	&	1.41\tablenotemark{a}	&	1.27	&	5.74:	&	6.15	\\
1562	&	M1III	&	\nodata	&	2.30	&	3.36	&	1.47	&	2.43	&	2.32	&	1.71	&	1.47	&	1.52	&	2.81	&	3.44	&	6.03	\\
HD36395 	&	M1.5V	&	\nodata	&	3.35	&	3.41\tablenotemark{b}	&	1.72	&	3.30:	&	3.04	&	1.97	&	2.64	&	0.98	&	2.46	&	5.12	&	8.48	\\
601	&	M2III	&	\nodata	&	1.86	&	8.13:	&	1.03	&	1.91	&	1.06	&	1.60	&	0.75	&	1.41\tablenotemark{a}	&	2.55:	&	2.22:	&	\nodata	\\
7696	&	M3III	&	\nodata	&	0.64	&	1.69	&	4.42:	&	0.67	&	1.07	&	2.87:	&	1.14	&	1.41\tablenotemark{a}	&	2.62:	&	1.66	&	1.38	\\
HD11469	 &	M3.5I	&	\nodata	&	3.98	&	2.21	&	1.96:	&	1.10	&	1.59	&	1.25:	&	1.41\tablenotemark{a}	&	1.33	&	1.65	&	1.41\tablenotemark{a}	&	0.37	\\
921	&	M4II	&	\nodata	&	1.70	&	2.40:	&	0.99	&	1.08	&	0.95	&	\nodata	&	1.41\tablenotemark{a}	&	0.69	&	3.15:	&	1.41\tablenotemark{a}	&	\nodata	\\
7139	&	M4III	&	\nodata	&	58.44:	&	8.86	&	3.37\tablenotemark{b}	&	2.80	&	4.59	&	3.01	&	3.72	&	1.04	&	5.20:	&	5.13	&	11.83	\\
9099	&	M4III	&	\nodata	&	3.49	&	3.99\tablenotemark{b}	&	1.79	&	2.93:\tablenotemark{b}	&	3.01\tablenotemark{b}	&	2.65\tablenotemark{b}	&	1.20\tablenotemark{b}	&	1.56	&	3.15\tablenotemark{b}	&	3.16\tablenotemark{b}	&	\nodata	\\
867	&	M6III	&	\nodata	&	8.07:	&	3.43:	&	1.91:	&	1.47:\tablenotemark{b}	&	1.99	&	2.63:	&	0.38	&	0.65	&	4.27:\tablenotemark{b}	&	3.50:\tablenotemark{b}	&	0.71	\\
9066	&	M7IIIe	&	\nodata	&	2.63	&	1.13\tablenotemark{b}	&	2.62	&	1.41\tablenotemark{a}	&	5.22:	&	2.27:\tablenotemark{b}	&	0.33\tablenotemark{b}	&	1.38	&	3.34	&	3.23\tablenotemark{b}	&	5.14	\\

\enddata																											
\tablenotetext{a}{Three sigma upper limit}																											
\tablenotetext{b}{Wavelength off by more than 8 \AA}																											
\tablenotetext{c}{The FWHM was fixed at 35 \AA~and the cored depth is reported rather than equivalent width.}																											
\end{deluxetable}

\clearpage

\begin{figure}
\figurenum{1}
\centering
\plotone{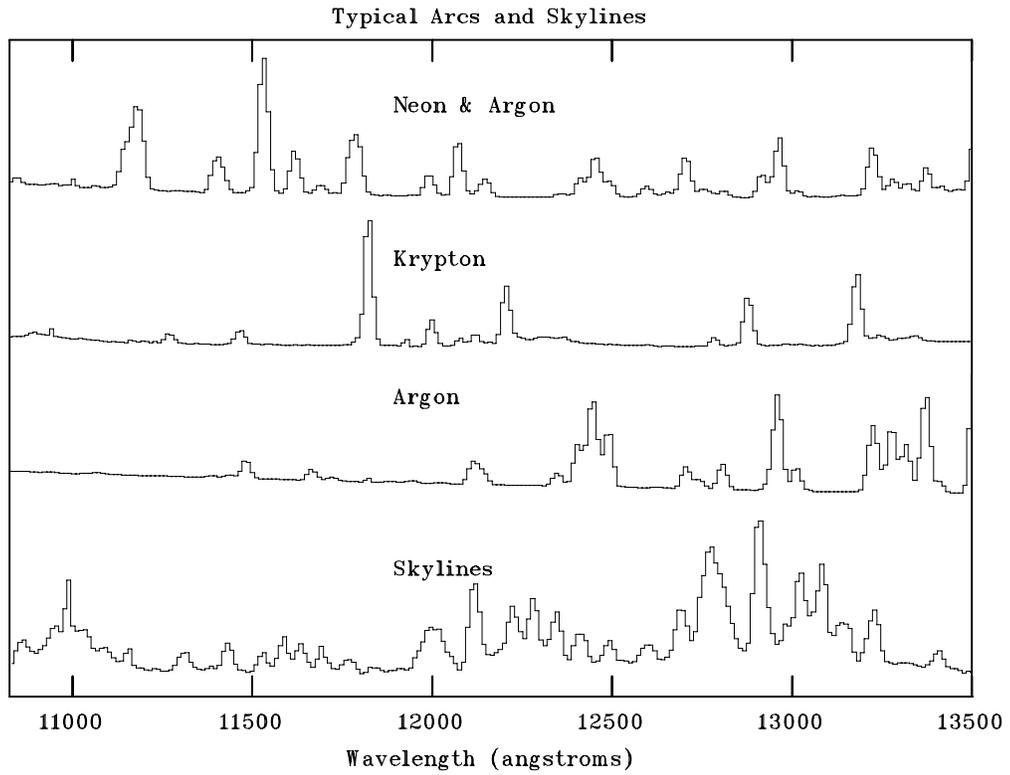}
\caption[]{Arc lamp used for calibrating data.  Sky lines were used when no arc lamp was available.  Each of the spectra have been scaled to unity and an arbitrary additive constant has been added to the fluxes on the vertical scale.}
\end{figure}

\clearpage

\begin{figure}
\figurenum{2}
\centering
\plotone{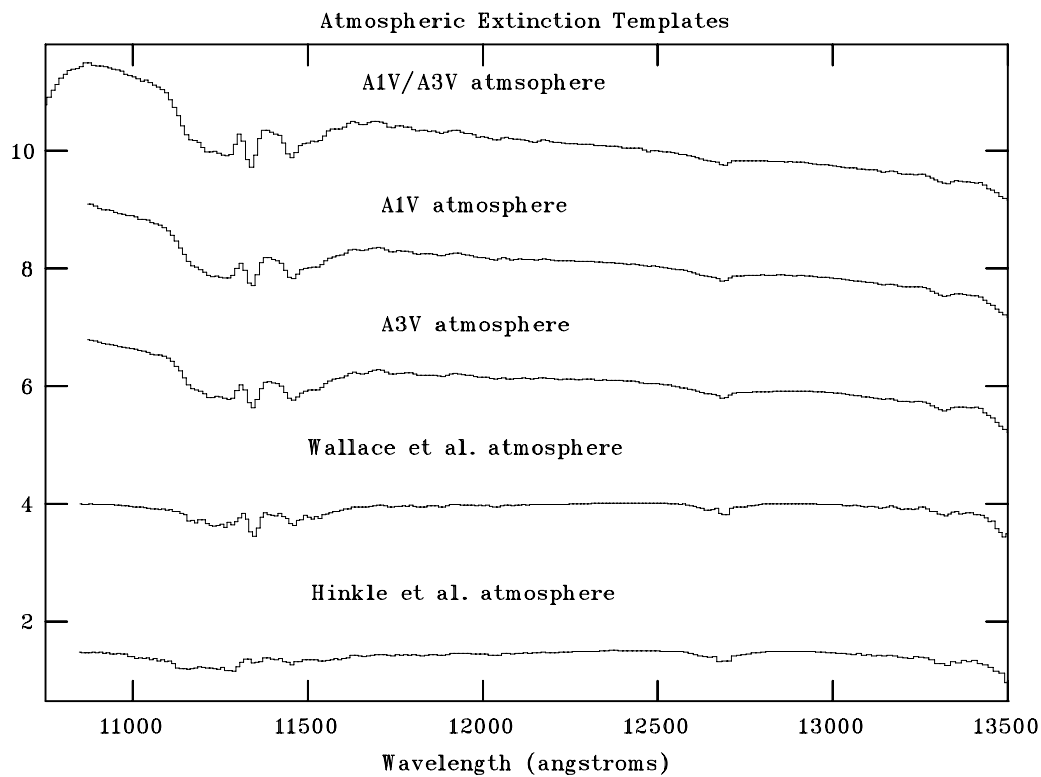}
\caption[]{Atmosphere extinction spectra.  The A1V/A3V atmosphere spectrum was used for the region between 1.07\mic~and 1.10\mic, while the A1V atmosphere spectrum was used for the region between 1.15\mic~and 1.35\mic.  The other three atmosphere spectra were also tried to check that the absorption features identified were not dependent on the atmosphere spectra used.  The atmospheric template used for atmosphere correction in the solar spectrum and Arcturus spectrum are those published by Wallace et al. (1996) and Hinkle et al. (1995), respectively.  The spectra have been scaled to unity and shifted vertically by 2.0.}
\end{figure}

\clearpage

\begin{figure}
\figurenum{3}
\centering
\caption[]{Blowup windows of each feature identified with spectra of several stars of similar spectral type(s) averaged together.  The spectral types included in the averaged spectrum are given to the right of the spectrum and the number of stars averaged together is in parentheses.  Also plotted are the high resolution solar and Arcturus spectra, labeled Sun(H) and Arct.(H) respectively, and the same spectra degraded to match our resolution, labeled Sun(L) and Arct.(L) respectively.  The horizontal lines indicate the window in which the labeled feature was measured.  The larger vertical dashed lines mark the centers of the features.  For those feature for which a wavelength shift was present, the smaller vertical dached line marks the center of the feature in hotter stars (earlier than K) and the larger dashed line is the center of the feature for cooler stars (K and M stars).  The features shown in each figure are as follows: (a) Paschen $\gamma$ (1.0942\mic) and CN index, (b) 1.16 \mic, 1.17 \mic~ and Mg I (1.1831 \mic), (c) 1.19 \mic, 1.20 \mic, Si I (1.2034 \mic), and 1.21 \mic, (d) 1.28 \mic, and (e) 1.31 \mic~and 1.33 \mic.  All spectra have been scaled to 2 and shifted by an arbitrary additive constant along the vertical scale.  The emission features seen in the Arcturus spectrum are due to poor sky subtraction, as are apparent absorption features with FWHM $<$ 0.1 \AA, which appear as spikes of one line width in our plots.}
\end{figure}

\begin{figure}
\figurenum{3}
\centering
\end{figure}

\begin{figure}
\figurenum{3}
\centering
\end{figure}

\clearpage

\begin{figure}
\figurenum{4}
\centering
\epsscale{0.5}
\plotone{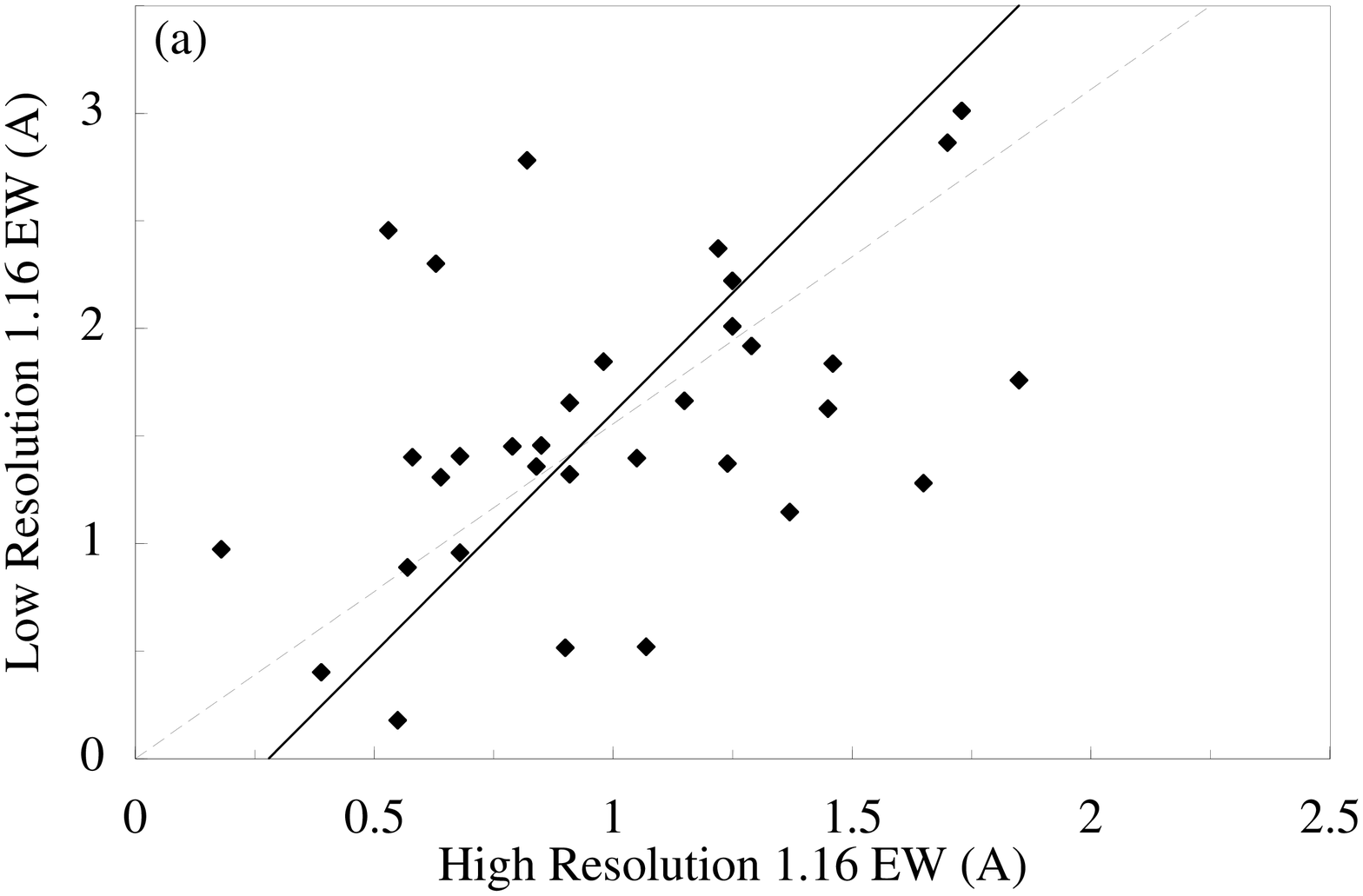}
\plotone{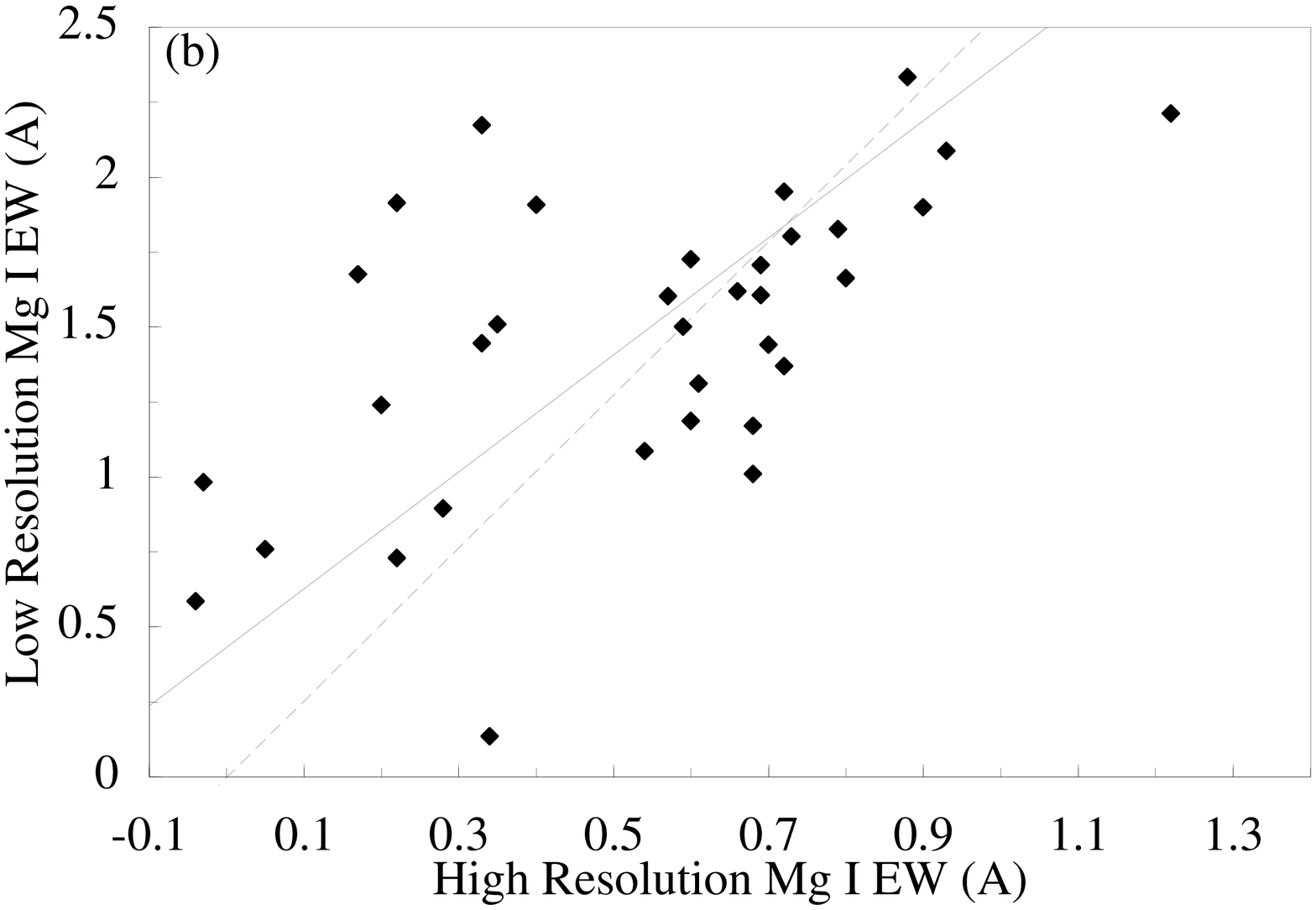}
\plotone{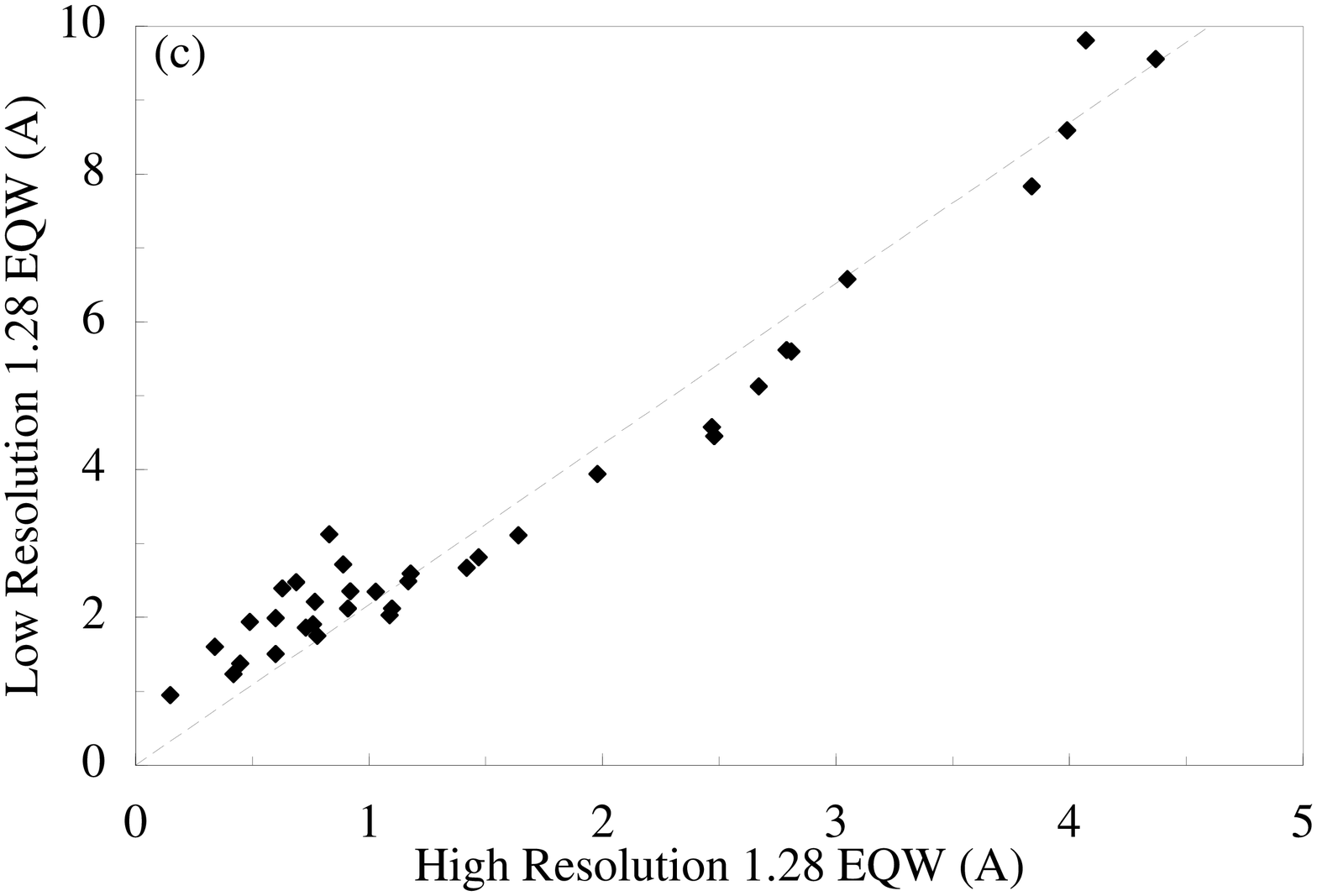}
\caption[]{High resolution equivalent width (W2000) versus the low resolution equivalent width measured using our technique for the (a) 1.16\mic, (b) Mg I, and (c) 1.28\mic~features.  The solid lines are the fits with no constraints on the y-axis intercept given by equations 1.1 and 1.3 for (a) and (b), respectively.  The dashed lines are the fits with a forced y-intercept of zero given by equations 1.2, 1.4, and 1.5 for (a), (b), and (c), repsectively.}
\end{figure}

\begin{figure}
\figurenum{5}
\epsscale{.7}
\plottwo{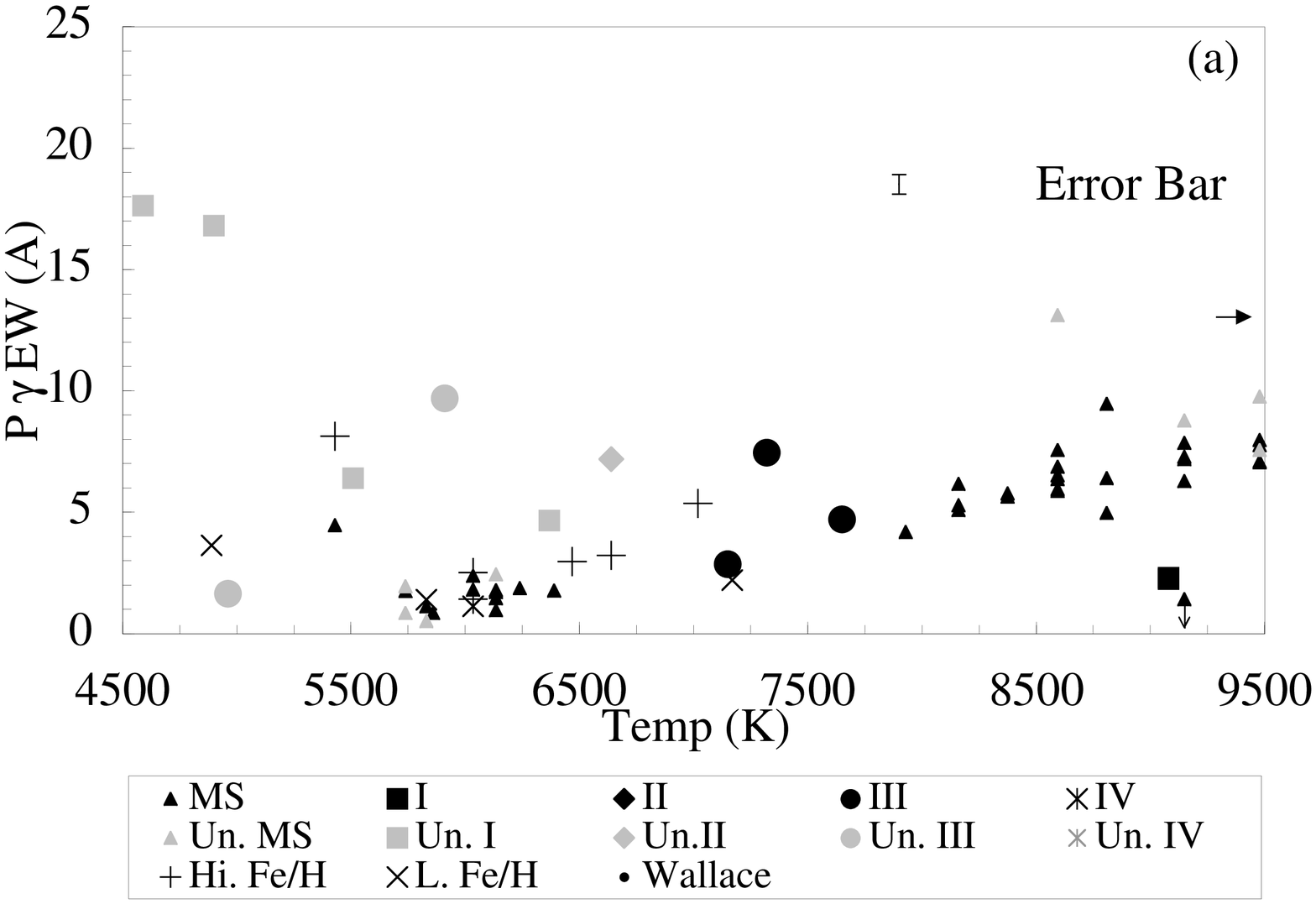}{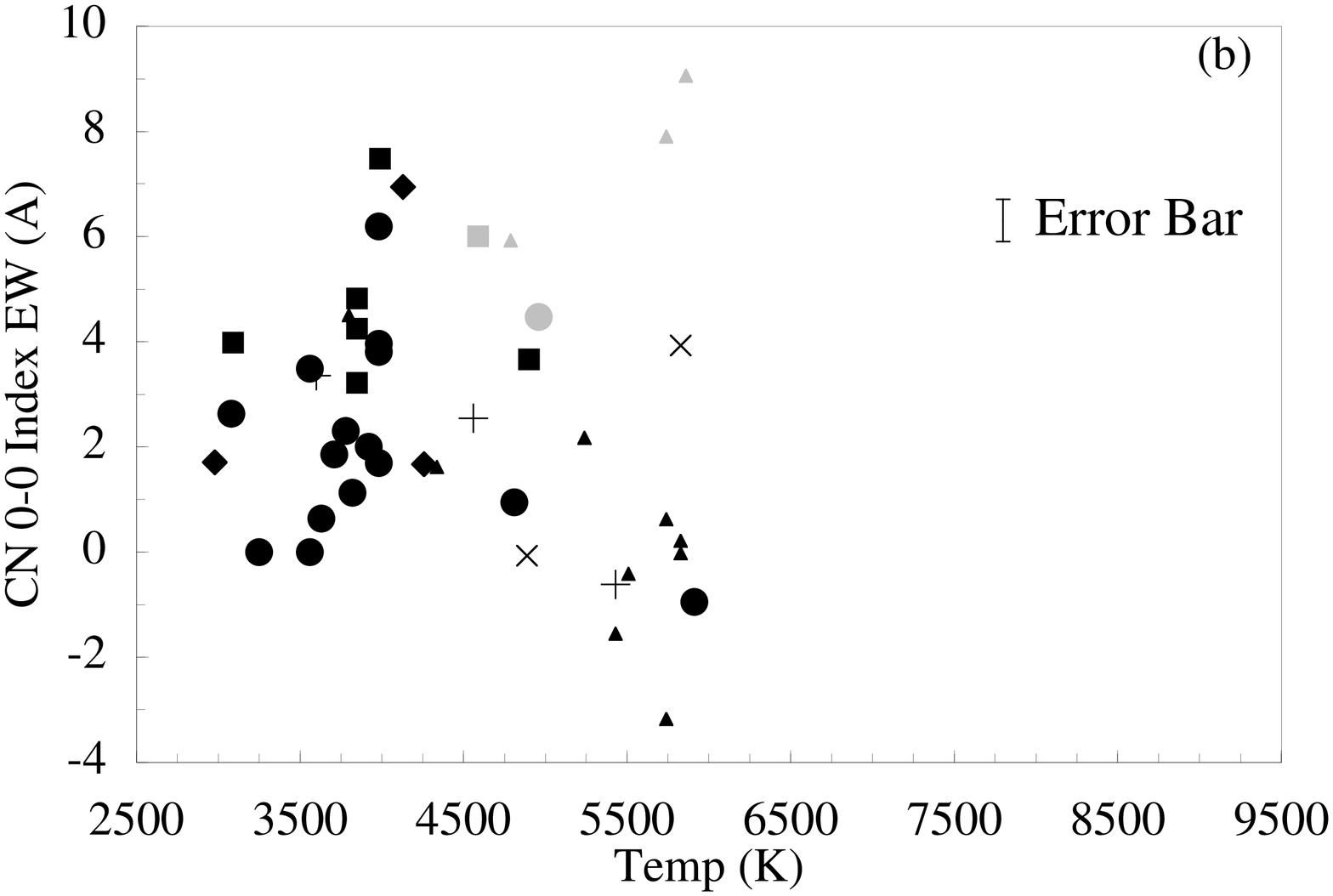}
\end{figure}

\begin{figure}
\figurenum{5}
\plottwo{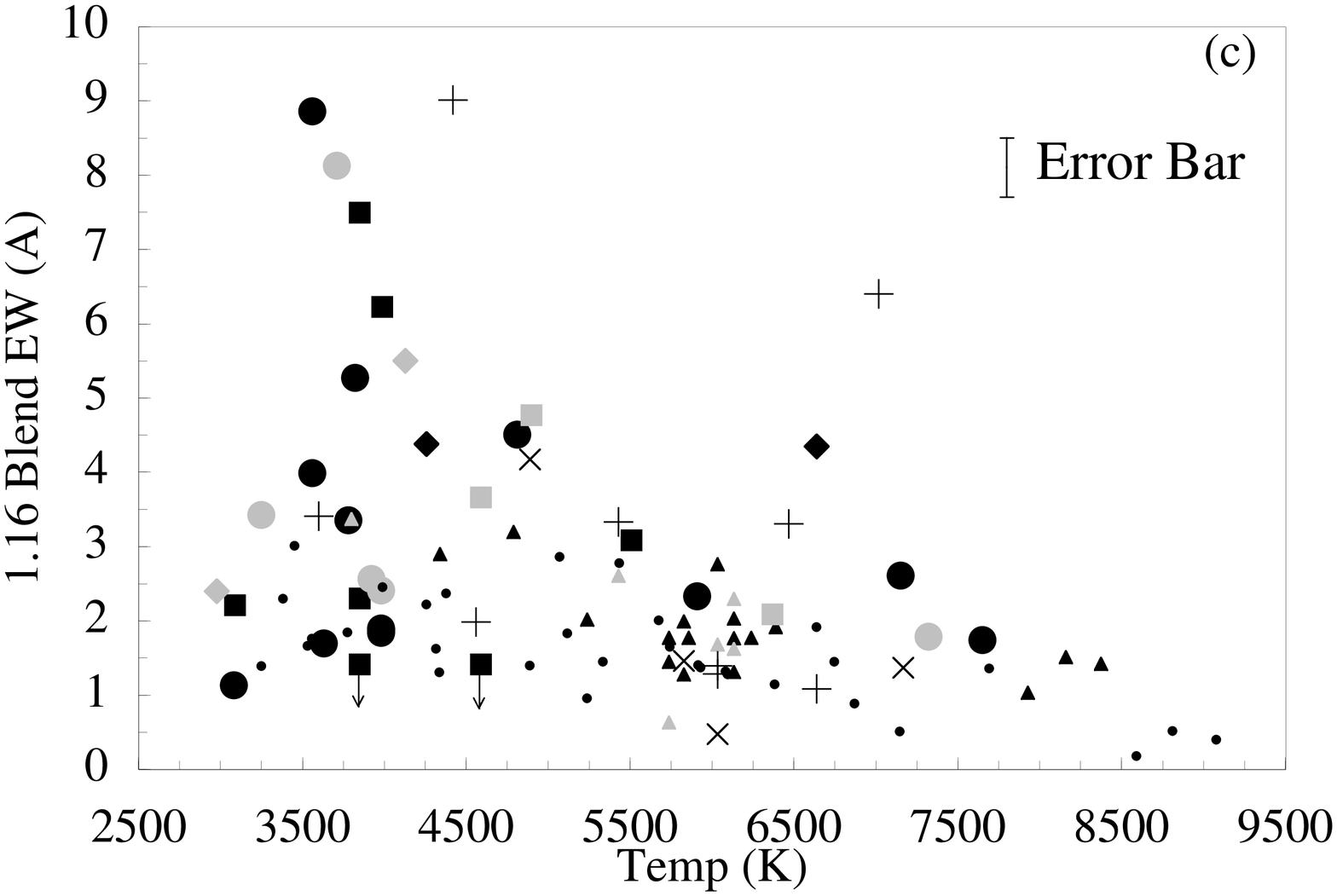}{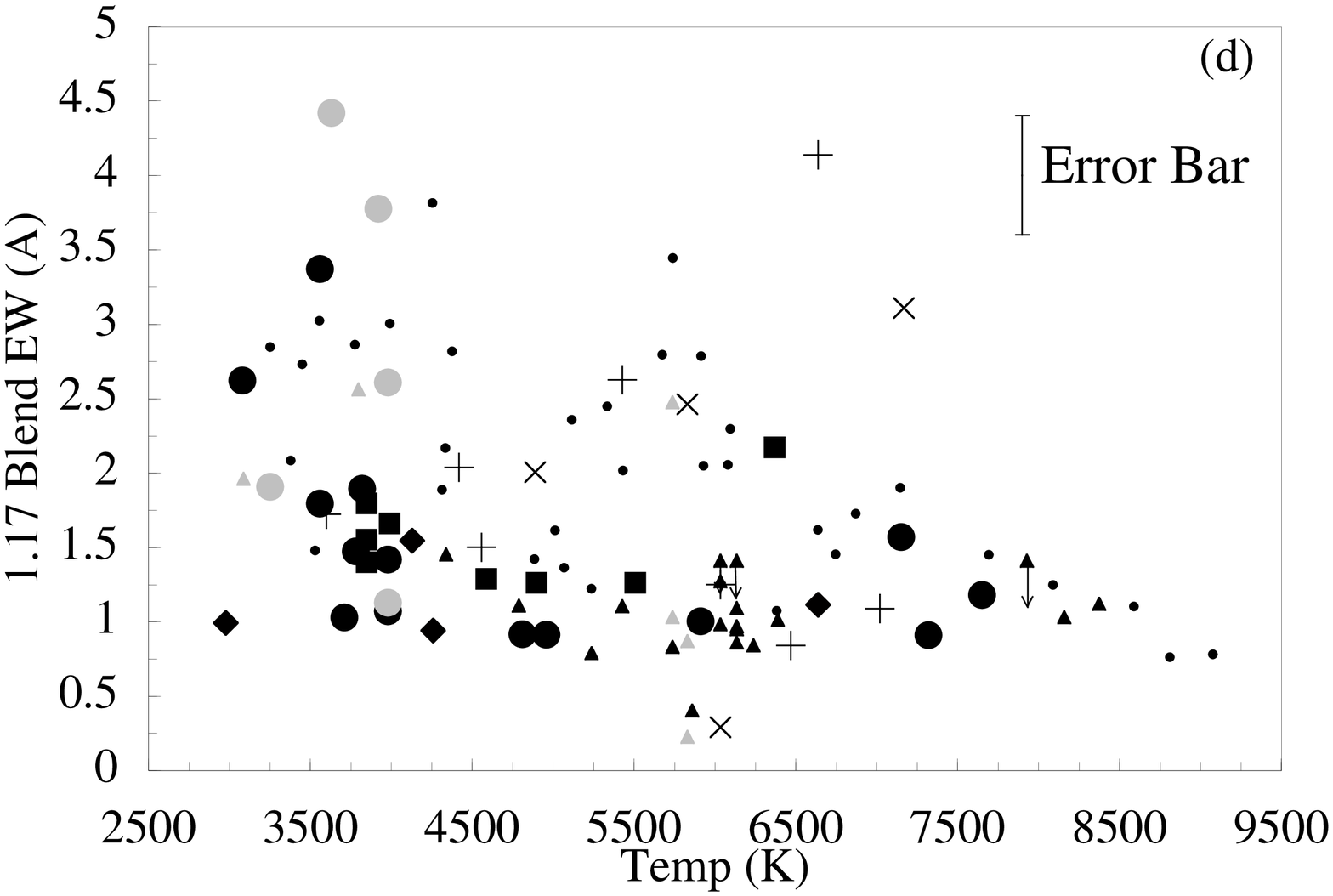}
\end{figure}

\begin{figure}
\figurenum{5}
\plottwo{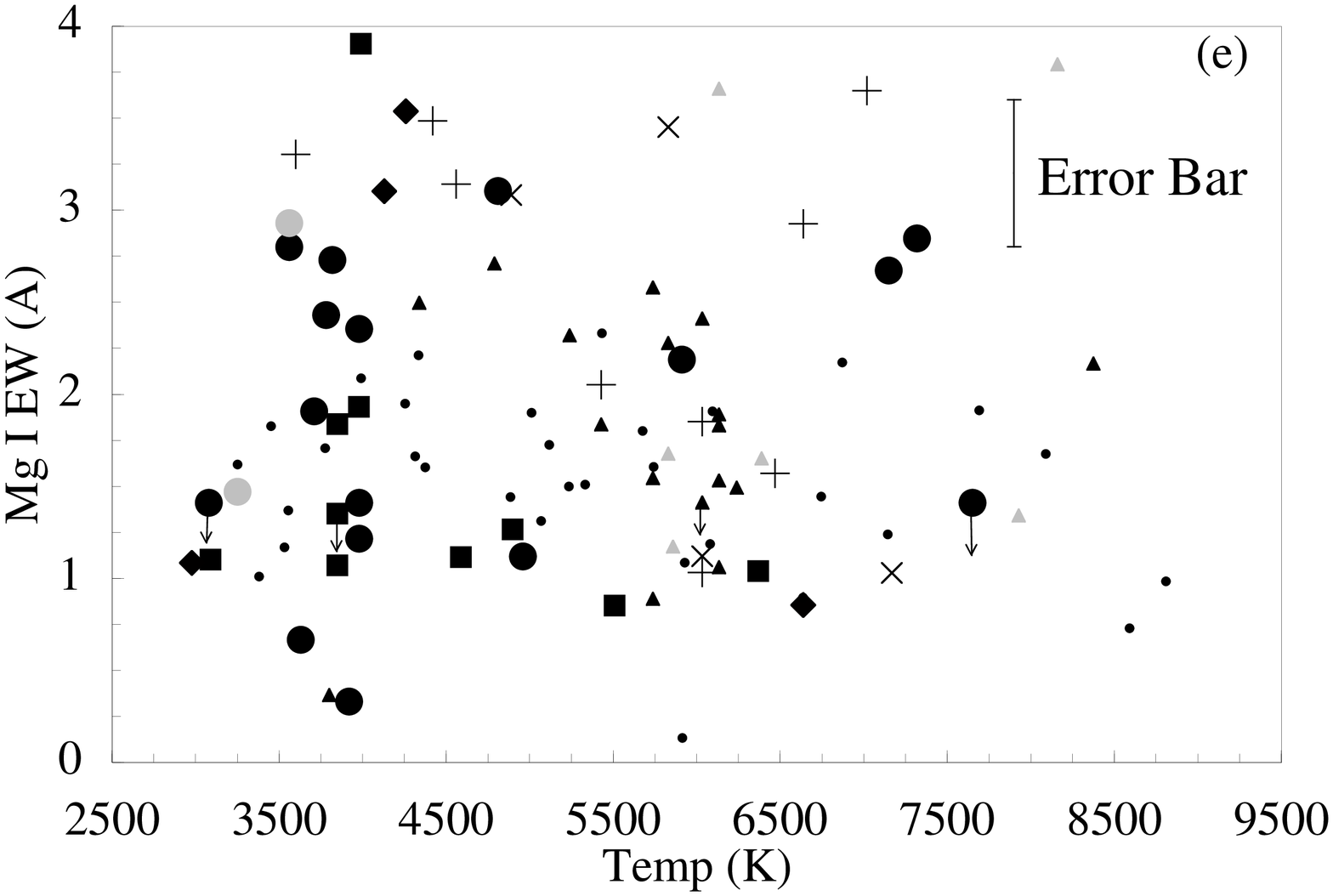}{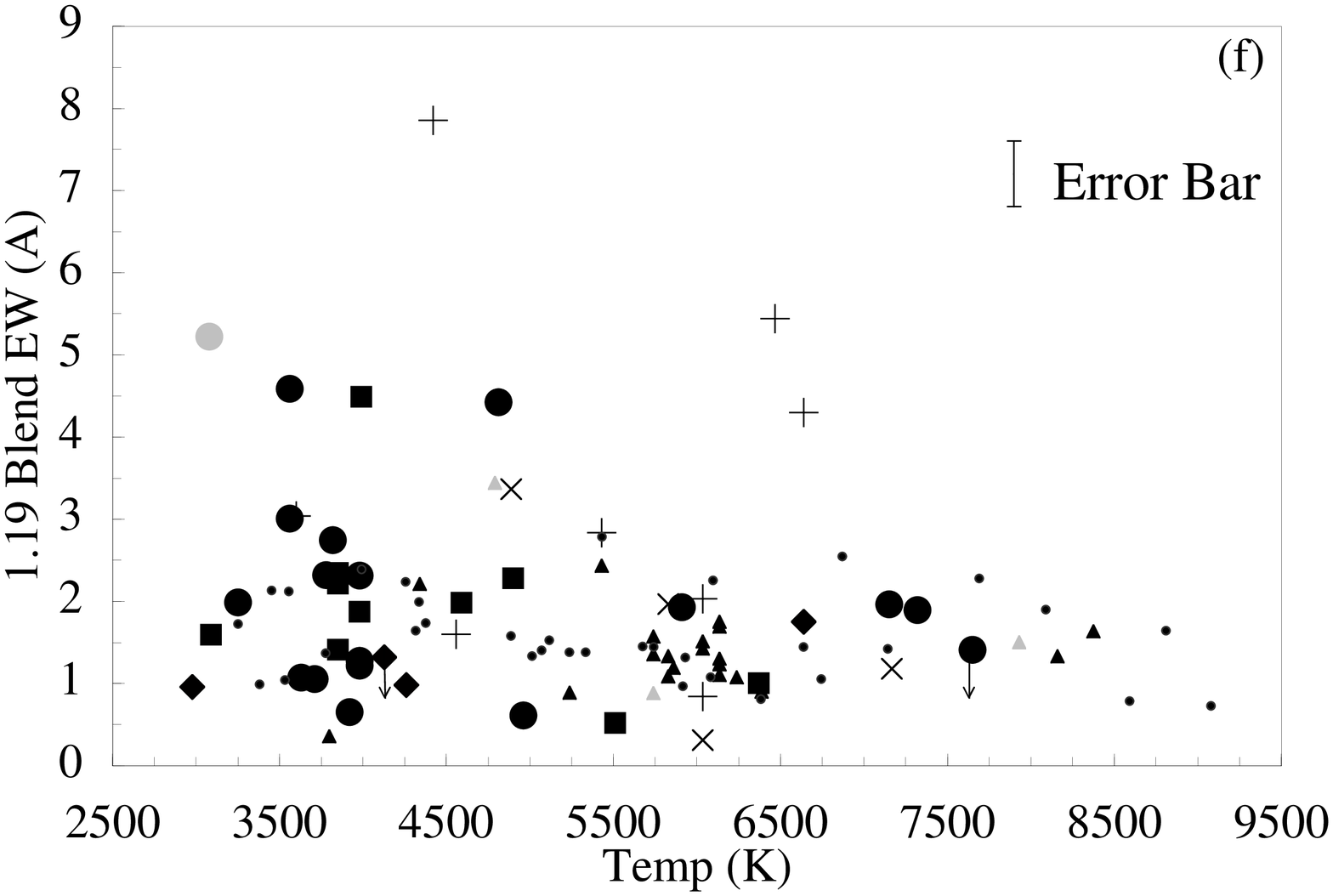}
\caption[]{Equivalent width versus temperature in degrees Kelvin for the (a) Paschen $\gamma$, (b) CN index, (c) 1.16\mic, (d) 1.17\mic, (e) Mg I, (f) 1.19\mic, (g) 1.20\mic, (h) Si I, (i) 1.21\mic, (j) 1.28\mic, (k) 1.31\mic, and (l) 1.33\mic~features. Solid black squares, diamonds, large circles, asterisks, and triangles represent luminosity class V, I, II, III, and IV respectively.  The same symbols in gray represent stars in the respective luminosity class with uncertain equivalent measurements ($\sigma$(EW) $>$ 0.4 \AA).  Crosses are stars with known high metallicity and ``x" represents stars known to have low metallicity.  All small circles are our measurements of spectra published by W2000 (see $\S$ 5).  An error bar is shown in the right corner to represent the equivalent width error that would be placed on each of the data points from our sample with a confident measurement (black symbols).  Horizontal arrows in (a) and (j) indicate more data exists at greater temperatures, but have not been included in the plot.}
\end{figure}

\clearpage
\begin{figure}
\figurenum{5}
\epsscale{.9}
\plottwo{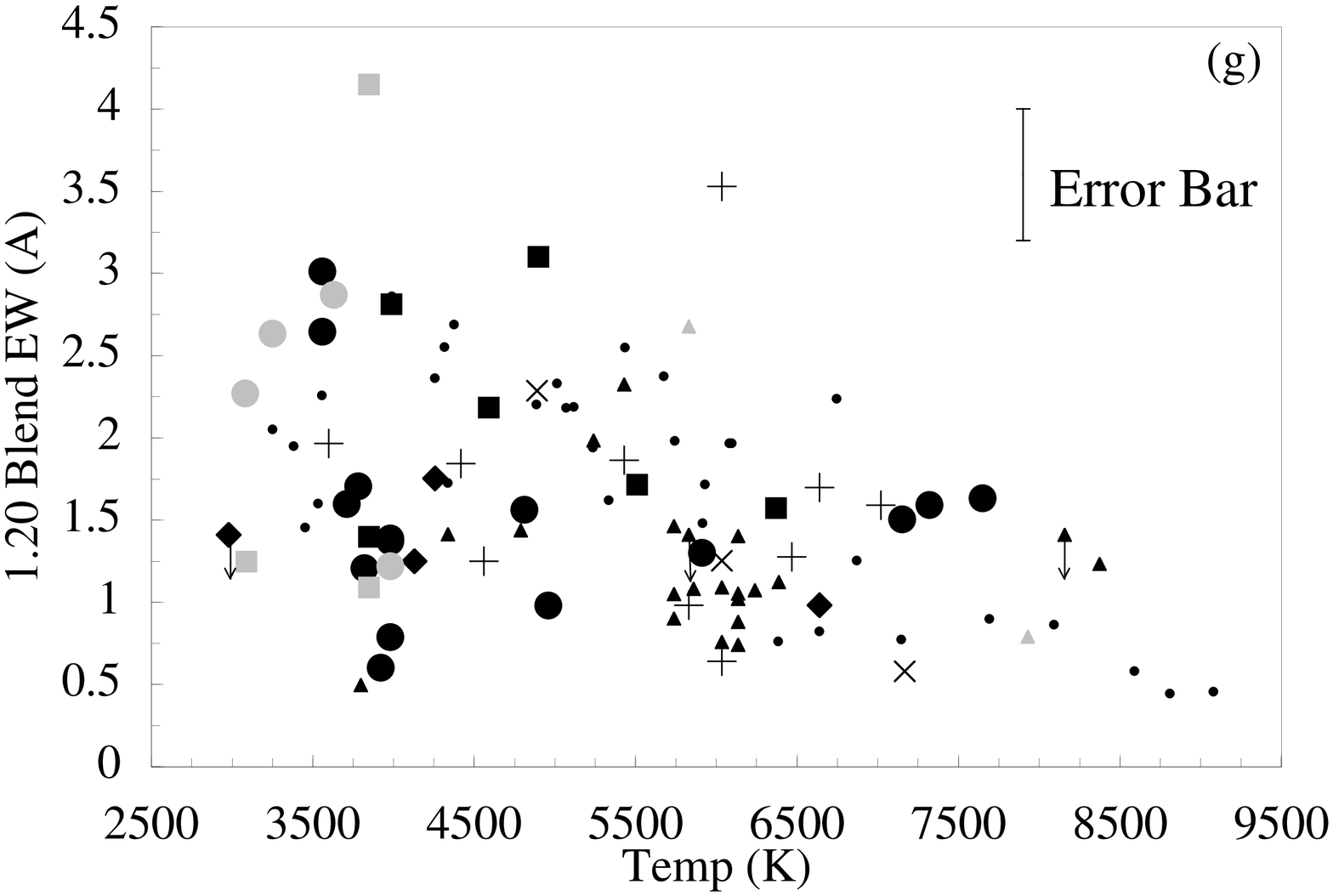}{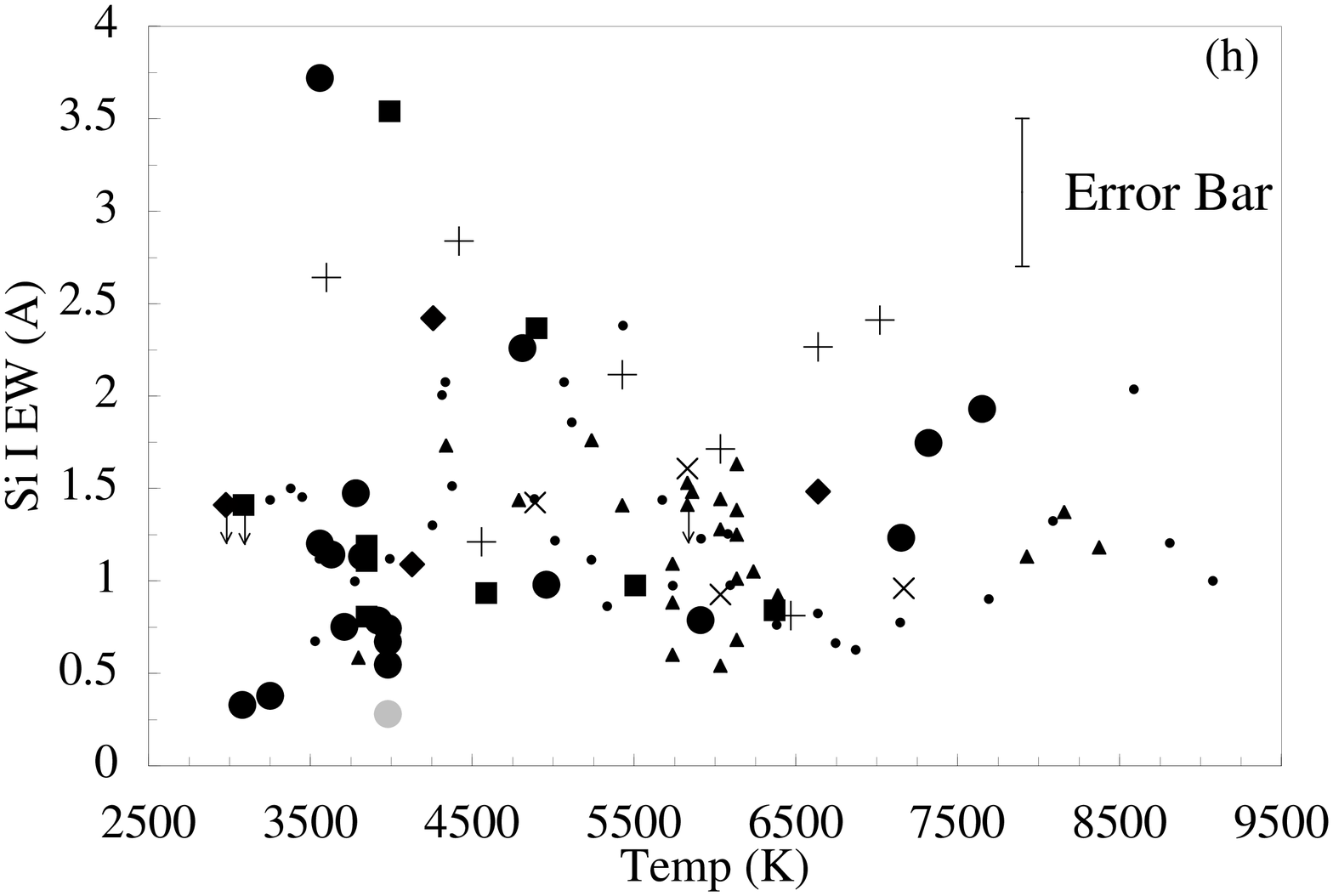}
\end{figure}

\begin{figure}
\figurenum{5}
\plottwo{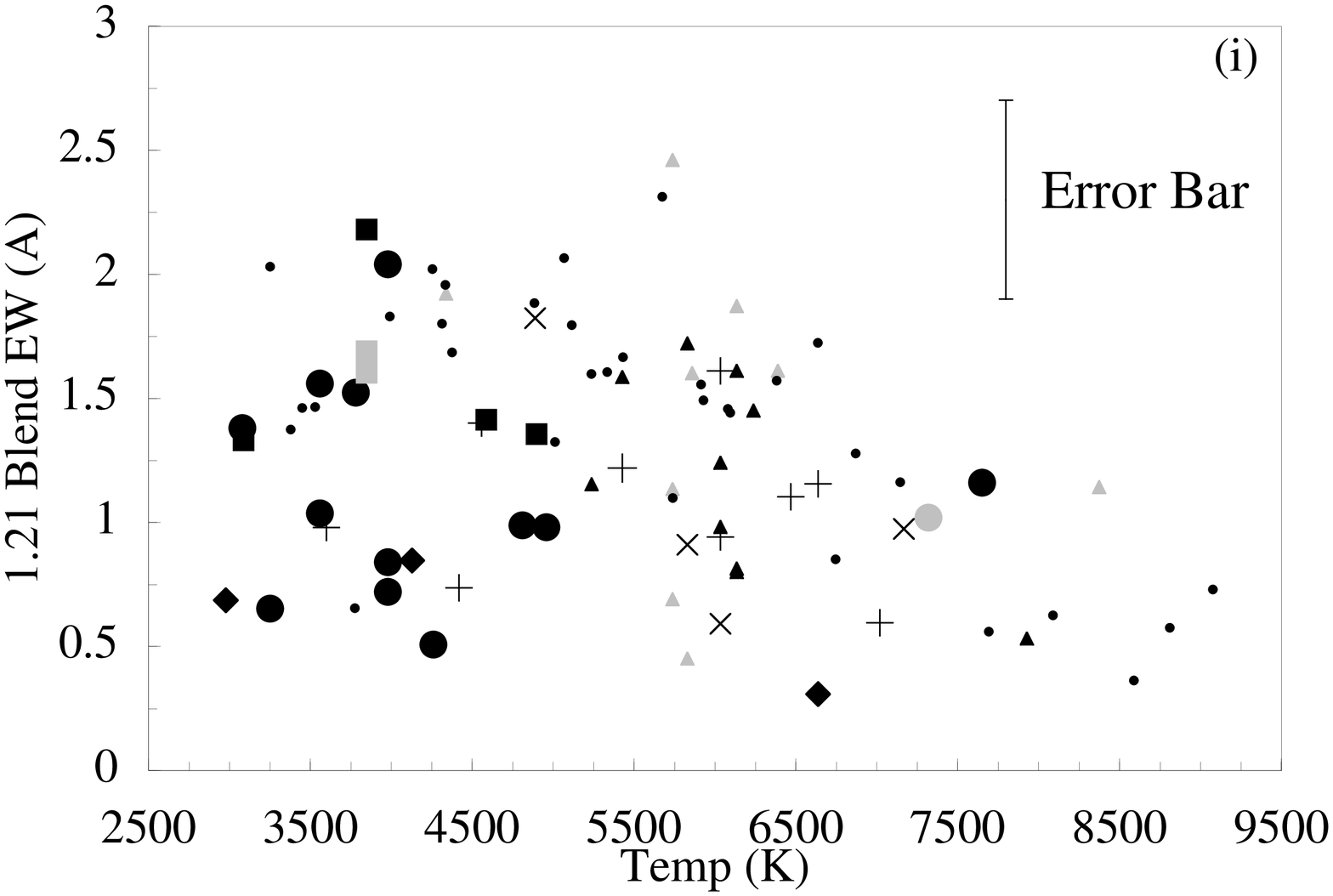}{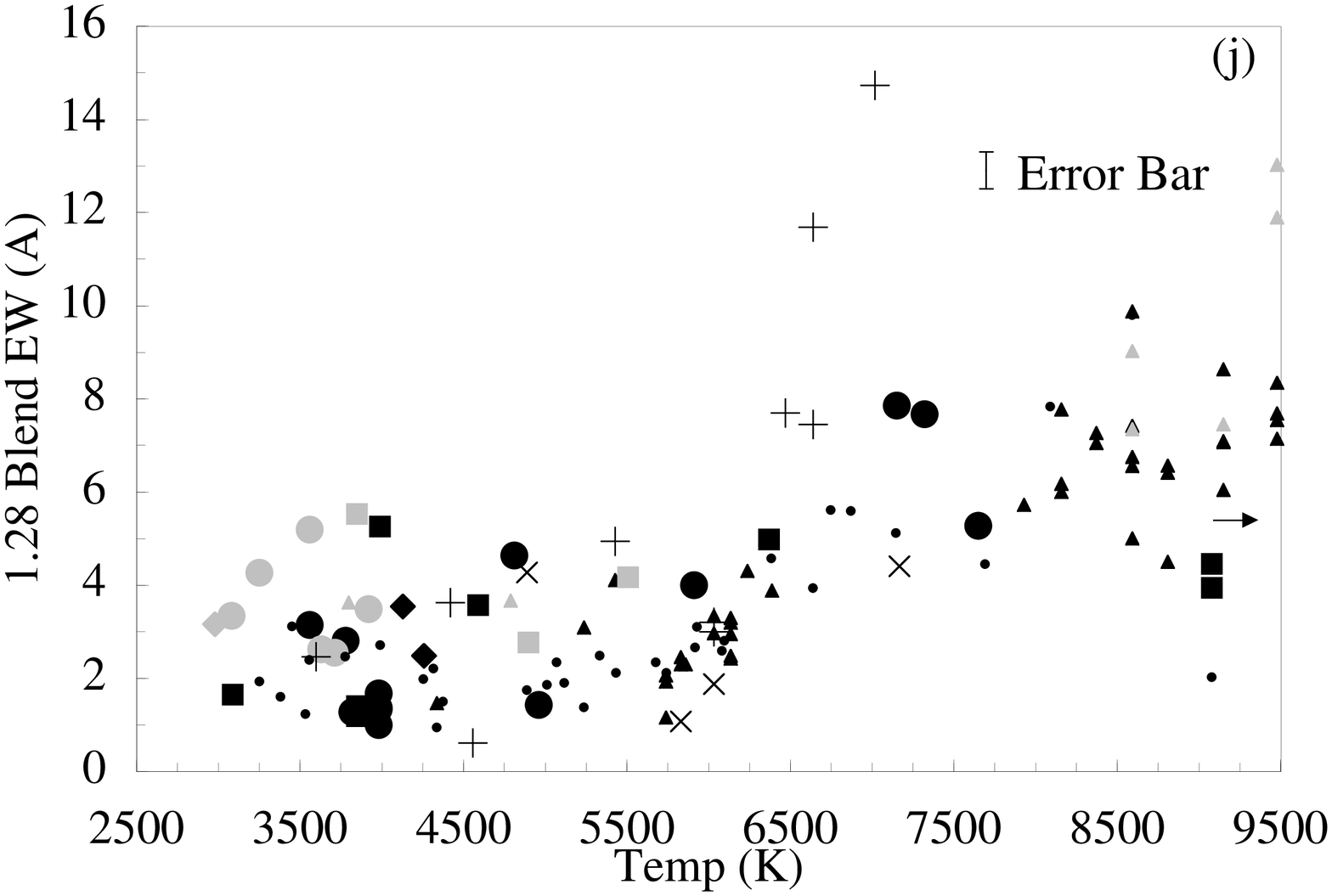}
\end{figure}

\begin{figure}
\figurenum{5}
\plottwo{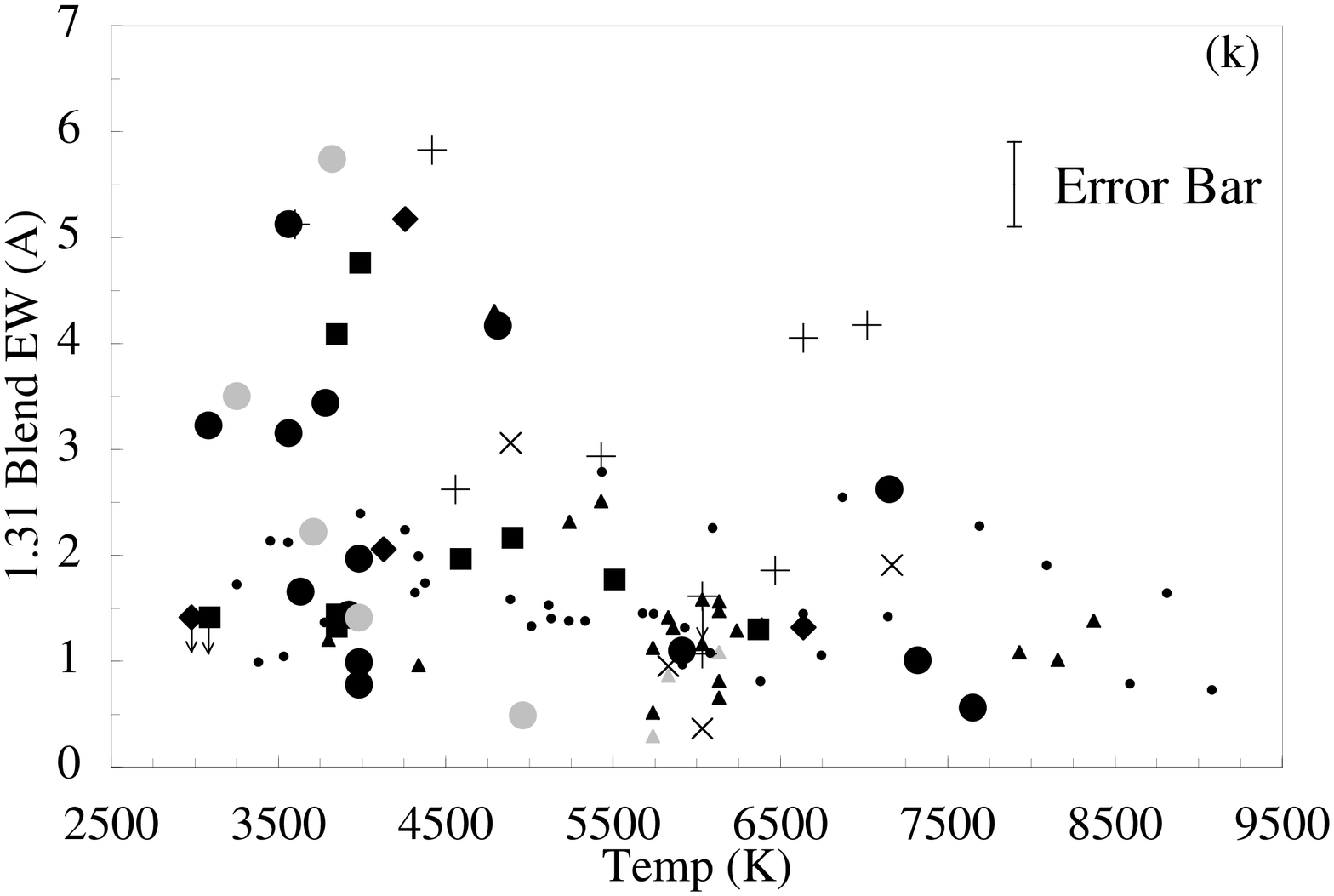}{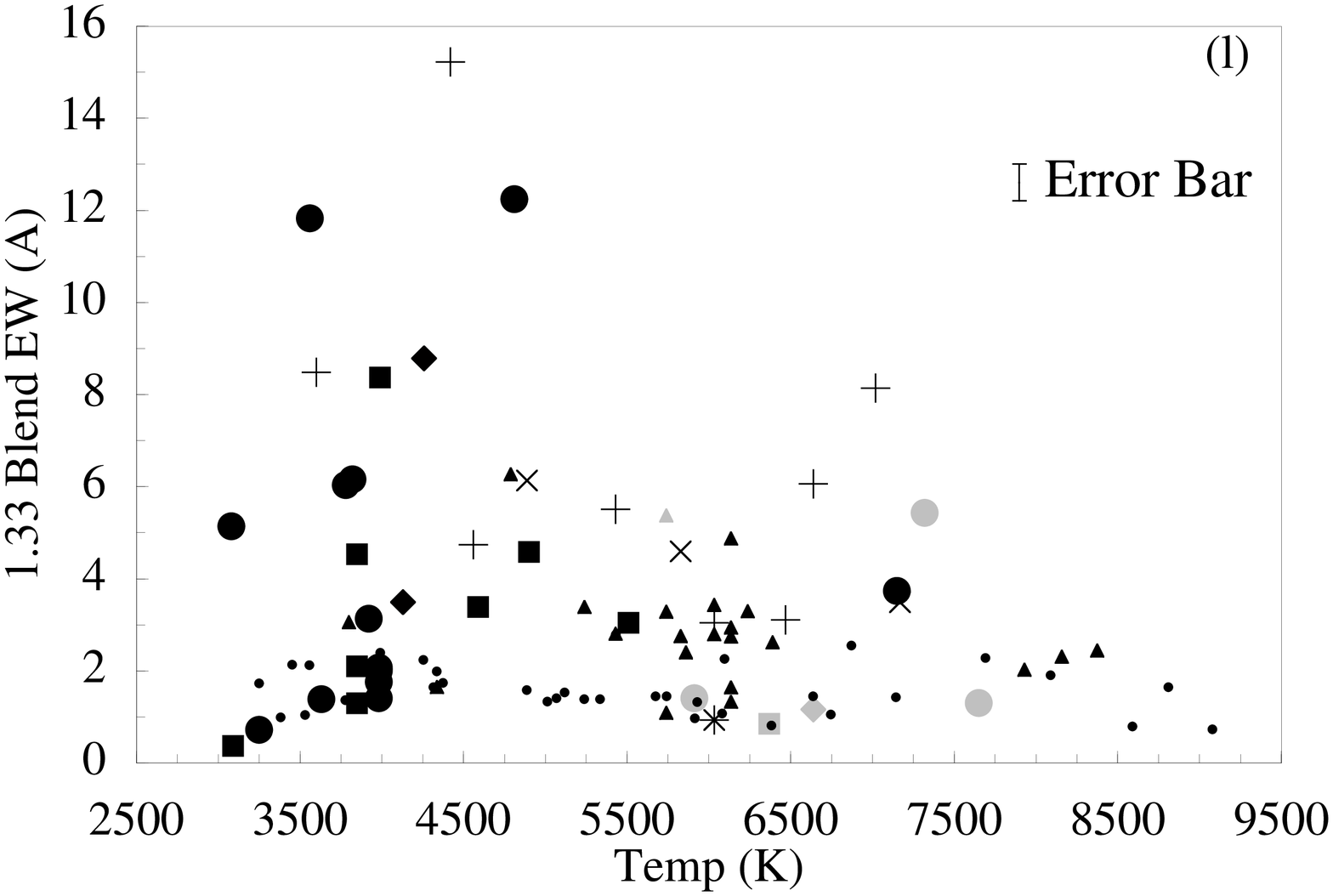}
\caption[]{- ${\it Continued}$}
\end{figure}

\clearpage

\begin{figure}
\figurenum{6}
\centering
\epsscale{0.8}
\plotone{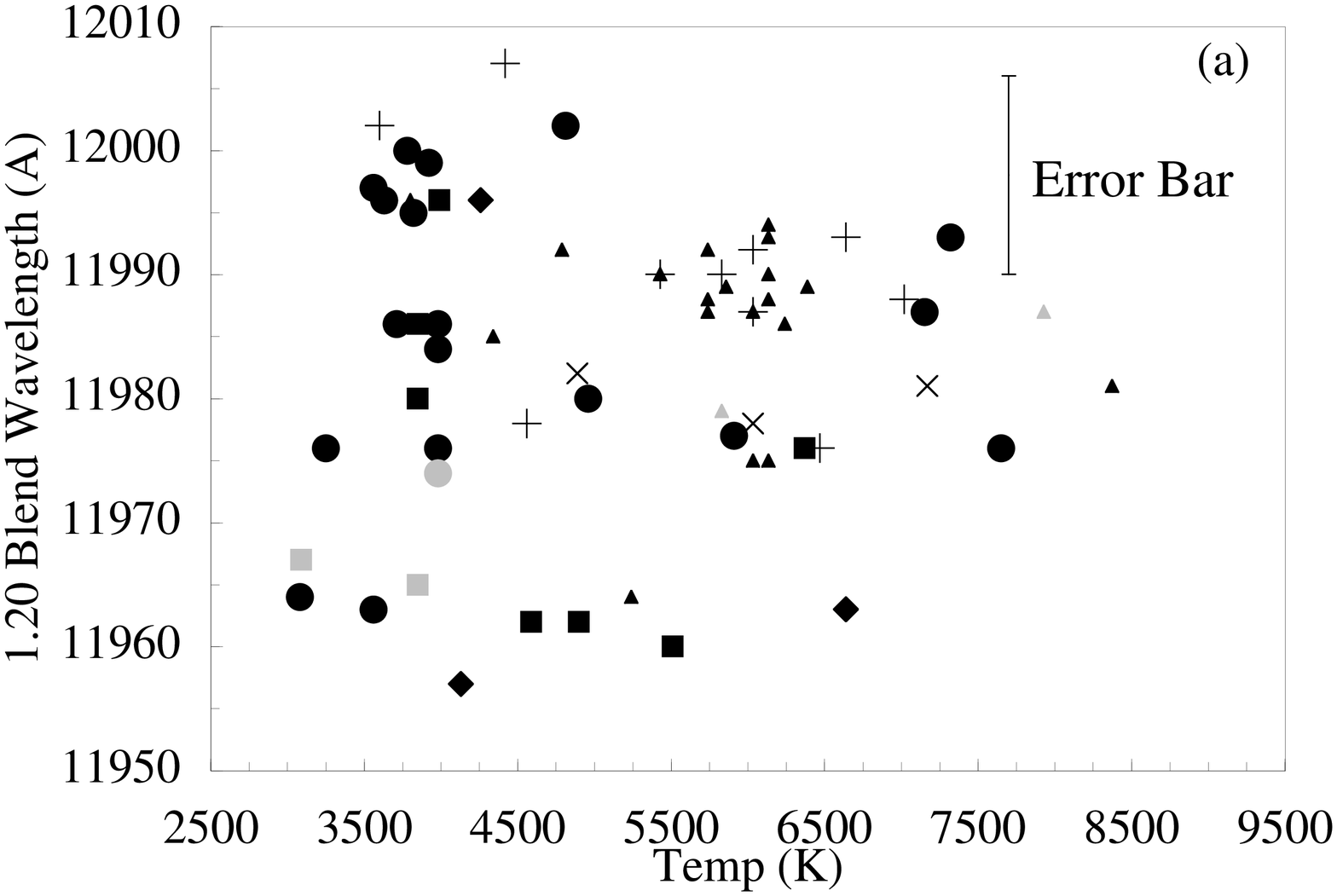}
\plotone{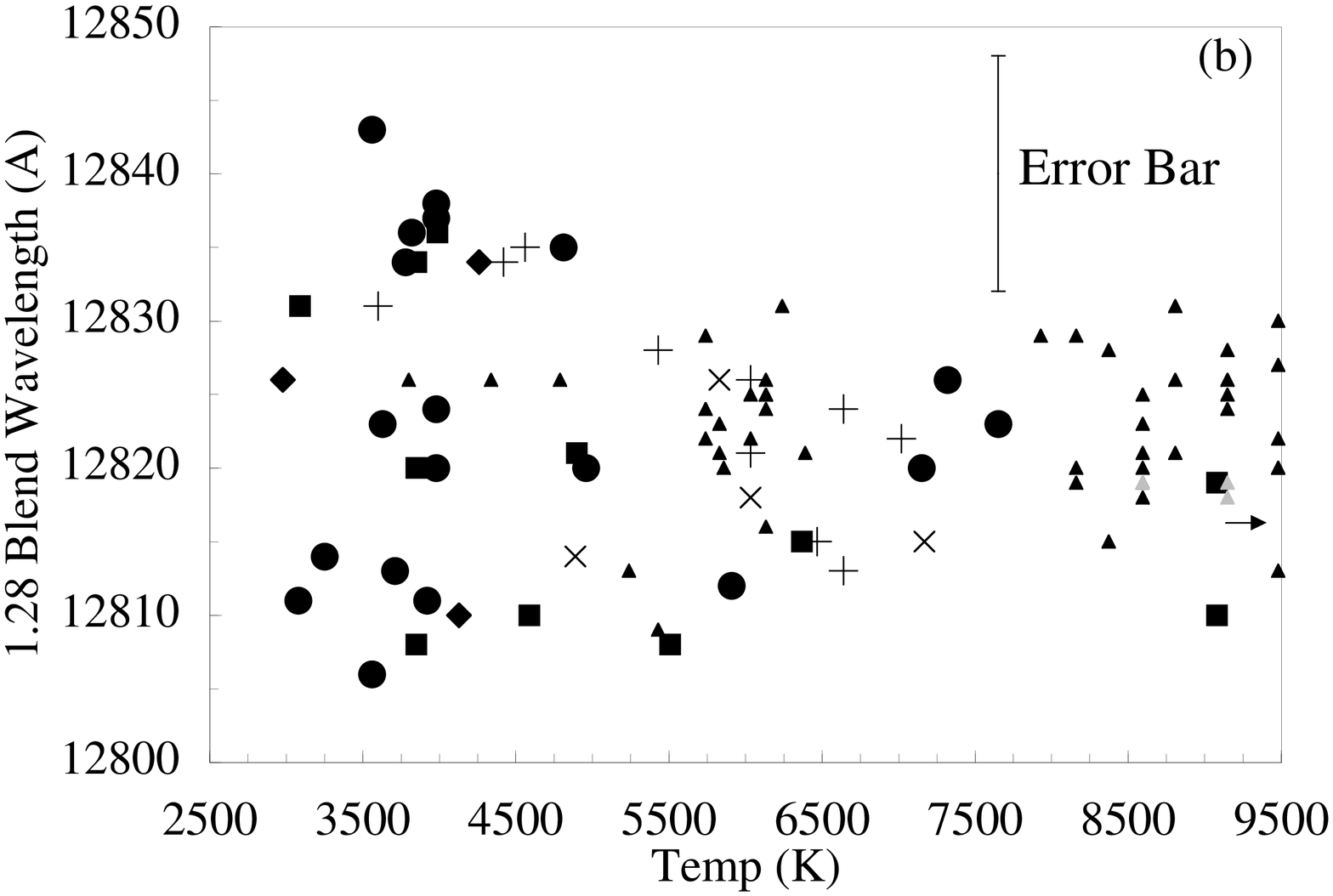}
\caption[]{Central wavelength versus temperature in degrees Kelvin for the (a) 1.20\mic~and (b) 1.28\mic~features.  See legend in Figure 5.  An error bar is shown in the right corner to represent the central wavelength error that would be placed on each of the data points from our sample with a confident measurement (black symbols).}
\end{figure}

\clearpage

\begin{figure}
\figurenum{7}
\centering
\epsscale{0.8}
\plotone{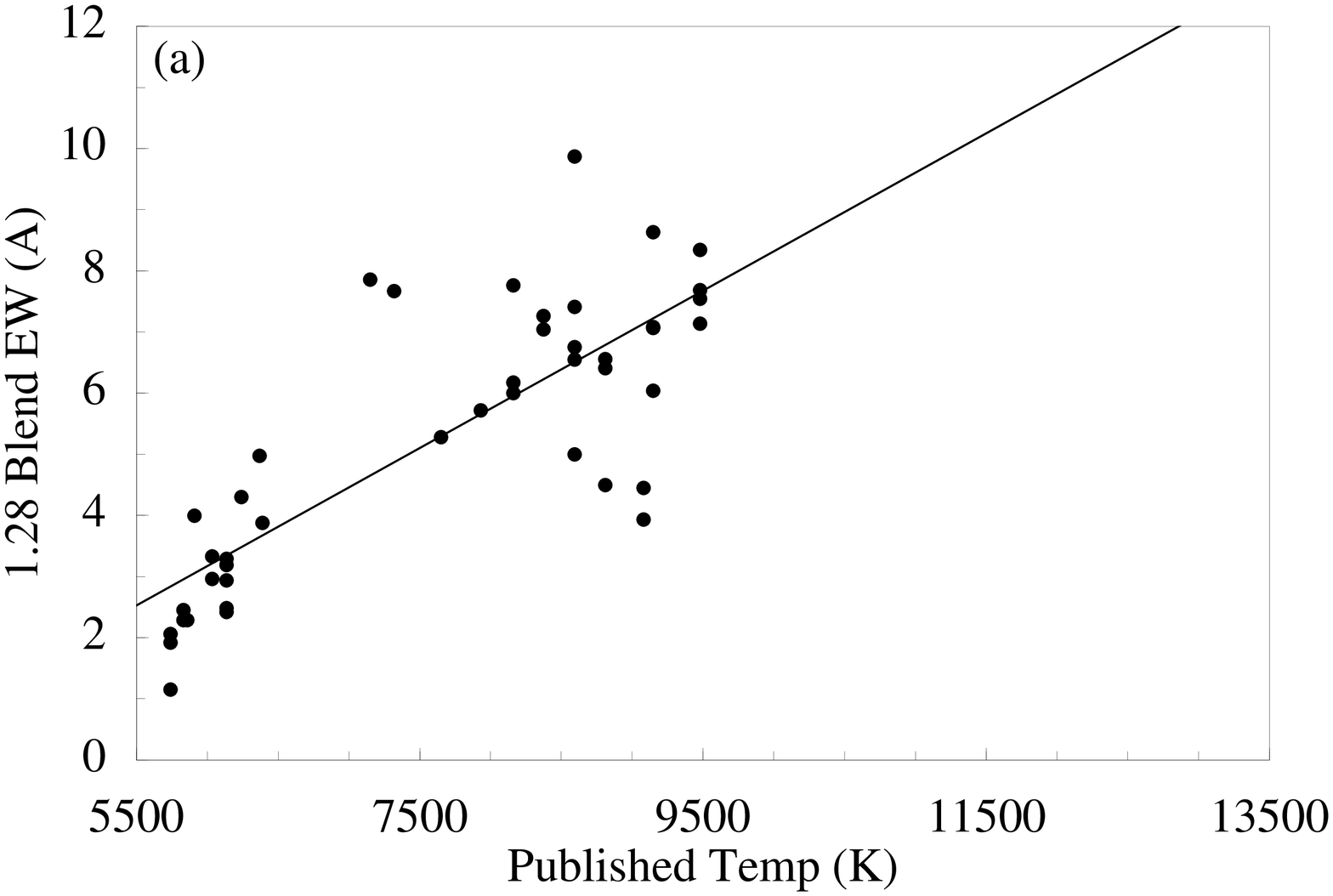}
\plotone{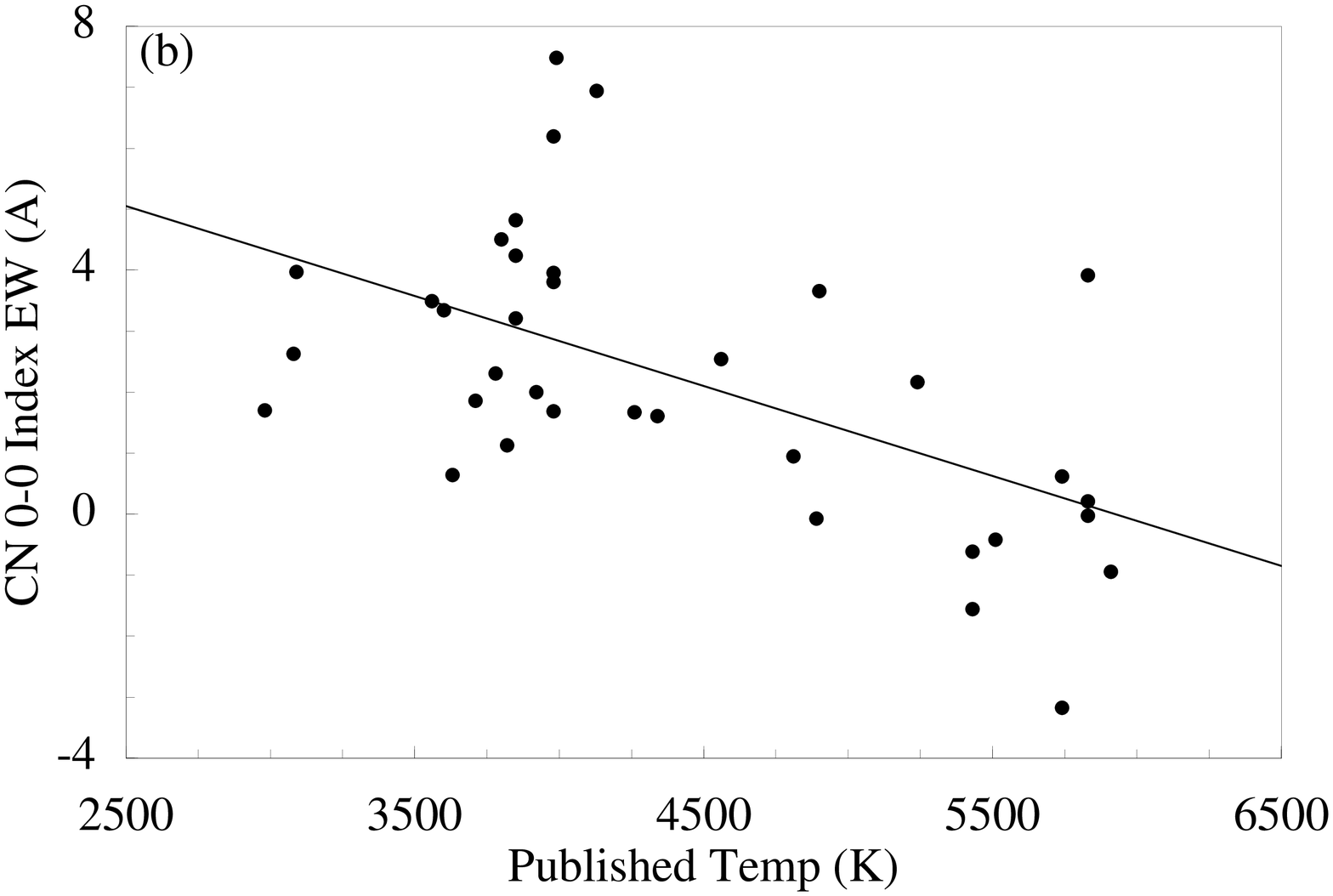}
\caption[]{Equivalent width versus the published temperature for all stars except those with uncertain equivalent width measurements ($\sigma$(EW) $>$ 0.4 \AA) or high/low metallicity (see $\S$ 5.3).  (a) 1.28\mic~blend for temperatures greater than 5500 K and less than 9500 K.  The solid line is the fit given by Equation 2.  (b) CN index for temperatures less than 6000 K.  The solid line is the fit given by Equation 3.}
\end{figure}

\clearpage

\begin{figure}
\figurenum{8}
\centering
\epsscale{1.0}
\plotone{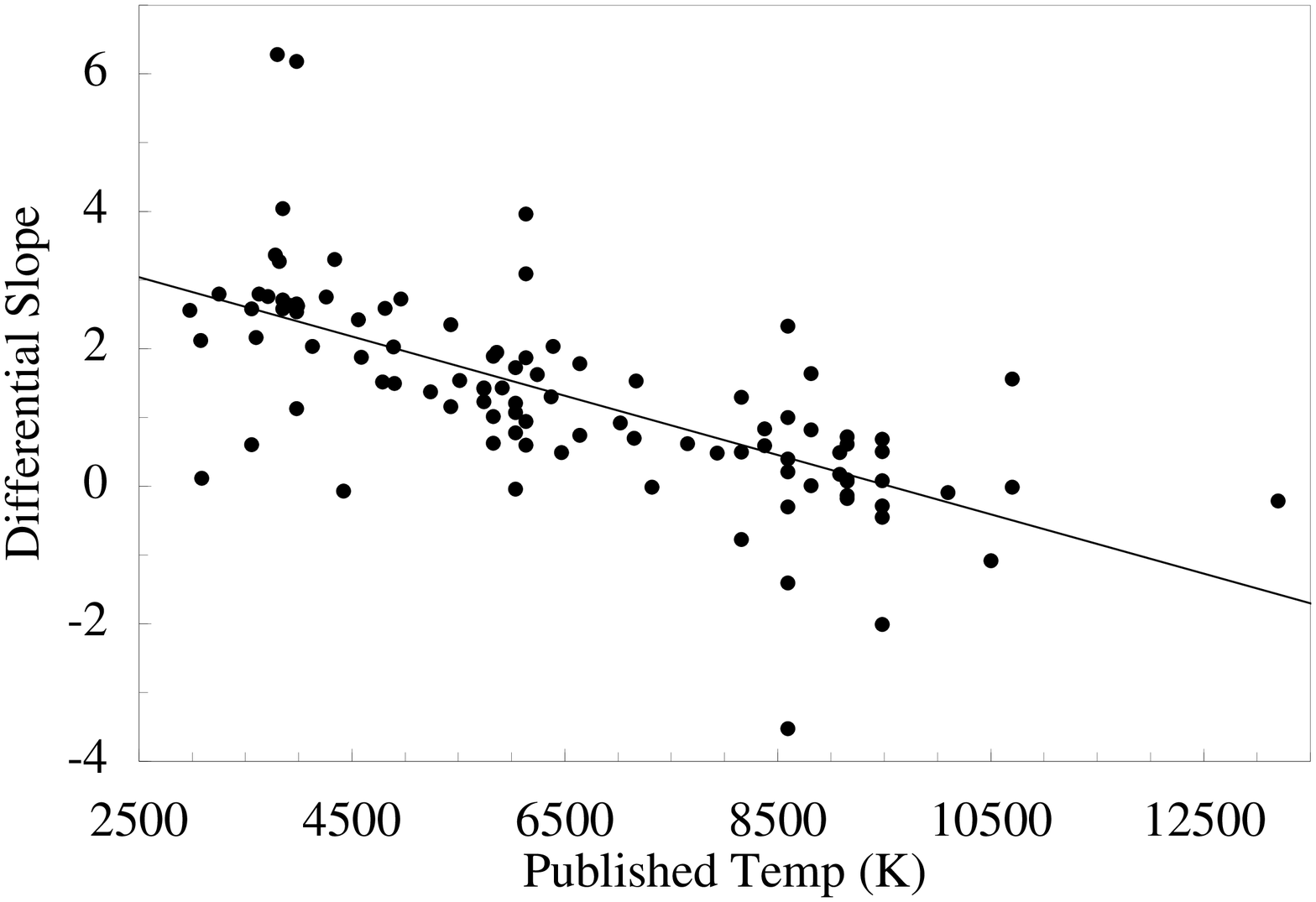}
\caption[]{Differential slope of the stellar spectra versus the published temperature for temperatures less than 13500 K.  The solid line is the fit given by equation 4.}
\end{figure}

\begin{figure}
\figurenum{9}
\centering
\epsscale{1.0}
\plotone{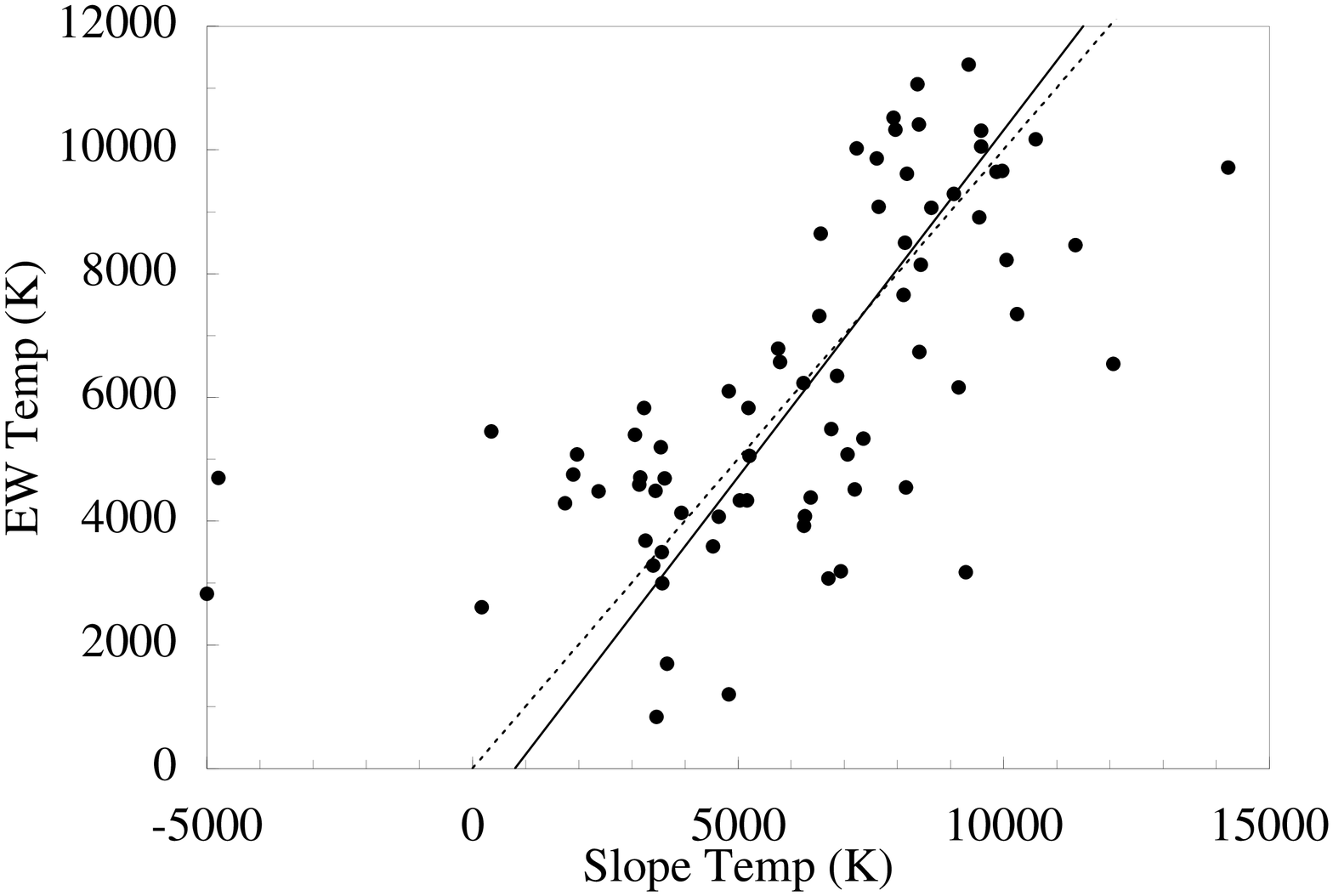}
\caption[]{Temperature determined from the equivalent width versus the temperature determined from the slope of the spectrum.  The solid line is the fit to the data and the dashed line is a one to one line.}
\end{figure}

\clearpage

\begin{figure}
\figurenum{10}
\centering
\epsscale{1.0}
\plotone{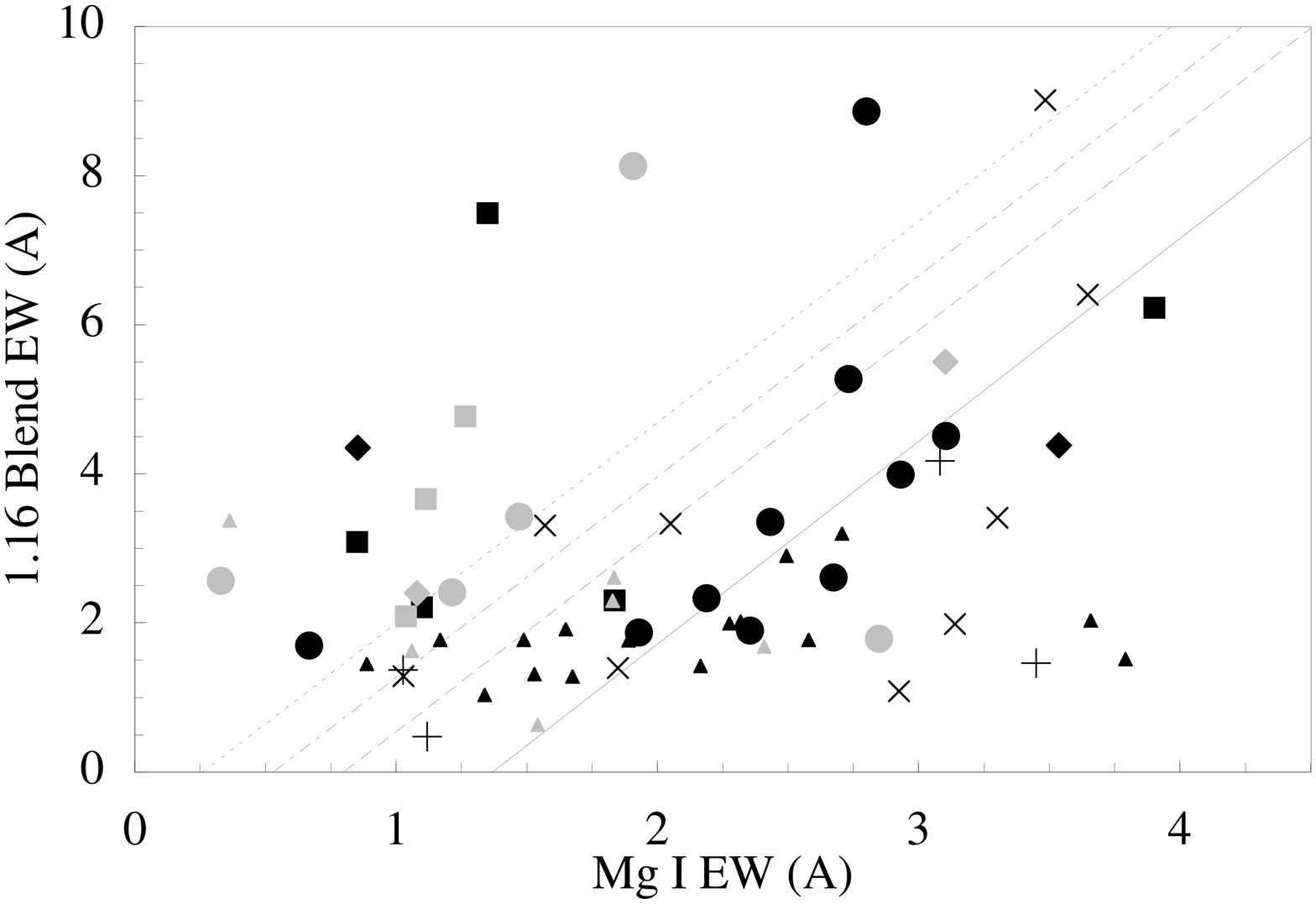}
\caption[]{Equivalent width of Mg I versus the equivalent width of 1.16\mic.  The solid line is a fit to the main sequence, the large dashed line is a fit to III, the medium dashed line is a fit to II, and the smallest dashed line is a fit to I luminosity class stars.  The fit is that given in equation 6.  See legend in Figure 5.}
\end{figure}

\end{document}